%% file: arXiv_v2.tex
\documentclass[aps,twocolumn,prl,superscriptaddress]{revtex4-1}

\usepackage{amsmath}
\usepackage{amsthm}
\usepackage{amssymb}
\usepackage{bm}
\usepackage{dsfont}
\usepackage{physics}
\usepackage[dvipsnames,rgb]{xcolor}
\usepackage{mathtools}
\usepackage{algcompatible}
\usepackage{newfloat}
\usepackage{hyperref}
\usepackage{cleveref}
\crefname{equation}{Eq.}{Equation}
\usepackage{soul}

\usepackage{pdfpages} 
\makeatletter
\AtBeginDocument{\let\LS@rot\@undefined}
\makeatother

\newcommand{\new}[1]{#1}

\usepackage{tikz}
\usetikzlibrary{shapes, calc, shapes, arrows, positioning}
\tikzstyle{neuron}=[draw,circle,minimum size=20pt,inner sep=0pt, fill=white]
\tikzstyle{stateTransition}=[thick]
\tikzstyle{learned}=[text=red]
\usepackage{standalone}

\usepackage[customcolors,shade]{hf-tikz} 

\definecolor{yucky}{HTML}{808000}
\definecolor{arylideyellow}{rgb}{0.91, 0.84, 0.42}
\definecolor{ballblue}{rgb}{0.13, 0.67, 0.8}

\definecolor{colup}{HTML}{008542}
\definecolor{colmid}{HTML}{FFE000}
\definecolor{coldown}{HTML}{E70001}

\newcommand{\liouv}[0]{\mathcal{L}}
\newcommand{\liouvd}[0]{\mathcal{L^\dagger}}
\newcommand{\orho}[0]{\hat{\rho}}
\newcommand{\oH}[0]{\hat{H}}

\newcommand{\tsigma}[0]{\tilde{\sigma}}
\newcommand{\vsigma}[0]{\bm{\sigma}}
\newcommand{\vtsigma}[0]{\bm{\tilde{\sigma}}}
\newcommand{\vv}[0]{\bm{v}}

\begin{document}
\title{Variational neural network ansatz for steady states in open quantum systems}
\author{Filippo Vicentini}
\affiliation{Laboratoire Mat\'{e}riaux et Ph\'{e}nom\`{e}nes Quantiques, Universit\'{e} Paris Diderot, CNRS-UMR7162, 75013 Paris, France}

\author{Alberto Biella}
\affiliation{Laboratoire Mat\'{e}riaux et Ph\'{e}nom\`{e}nes Quantiques, Universit\'{e} Paris Diderot, CNRS-UMR7162, 75013 Paris, France}

\author{Nicolas Regnault}
\affiliation{Laboratoire de Physique de l'Ecole Normale Sup\'{e}rieure, ENS, Universit\'{e} PSL, CNRS, Sorbonne Universit\'{e}, Universit\'{e} Paris-Diderot, Sorbonne Paris Cit\'{e}, Paris, France}

\author{Cristiano Ciuti}
\affiliation{Laboratoire Mat\'{e}riaux et Ph\'{e}nom\`{e}nes Quantiques, Universit\'{e} Paris Diderot, CNRS-UMR7162, 75013 Paris, France}

\date{\today}
\begin{abstract}
We present a general variational approach to determine the steady state of open quantum lattice systems via a neural network approach.
The steady-state density matrix of the lattice system is constructed via a purified neural network ansatz in an extended Hilbert space with ancillary degrees of freedom.
The variational minimization of cost functions associated to the master equation can be performed using a Markov chain Monte Carlo sampling. As a first application and proof-of-principle, we apply the method to the dissipative quantum transverse Ising model. \end{abstract}

\maketitle

In spite of the tremendous experimental progress in the isolation of quantum systems, a finite coupling to the environment \cite{Breuer2007} is unavoidable   and certainly plays a crucial role in the practical implementation of quantum information and quantum simulation protocols \cite{Devoret2013}. Moreover, through an active control of the environment via the so-called reservoir engineering, an open quantum manybody system can be prepared in non-trivial phases \cite{CarusottoRMP13,Noh2016,Hartmann2016} with also possible quantum applications \cite{Verstraete:2009fk,Barreiro:2011fj}.
The theoretical description of open quantum manybody systems is in general out-of-the equilibrium and much less developed than for equilibrium systems. A mixed state with a finite entropy can be described by a density matrix, whose evolution is described by a master equation. 
Recently, a few theoretical methods have been developed to solve the master equation of open quantum manybody systems, including analytical approaches based on the Keldysh formalism \cite{Sieberer2016,Maghrebi16}, numerical algorithms based on matrix product operator and tensor-network techniques \cite{MascarenhasPRA2015,CuiPRL2015,Jaschke2019,WernerPRL2016,OrusNat17}, cluster mean-field methods \cite{BiellaNLCE2018,Jin16}, corner-space renormalization \cite{FinazziPRL15,Rota17,RotaArx2018}, Gutzwiller mean-field \cite{Wim2018}, full configuration-interaction Monte Carlo \cite{NagyPRA18}, permutation-invariant solvers \cite{ShammahPRA2018} or efficient stochastic unravelings for disordered systems \cite{Vicentini18ArXiV}. The research in the field is very active, since the different methods are optimal for different specific regimes.  For example, the corner-space renormalization method is best suited for systems with moderate entropy, while matrix product operator techniques  to systems with short-range quantum correlations. 

In the last decade, the field of artificial neural networks has enjoyed a dramatic expansion and success thanks to remarkable applications in the recognition of complex patterns such as visual images or human speech (for a recent review see, e.g., \cite{LeCun2015}). The optimization (supervised learning) of the network is obtained by tuning  the weights quantifying the connections between neural units via a variational minimization of a properly defined cost function. The wavefunction of a manybody system is in general a complex quantity, which is hard to be recognized. Recent works have proposed to exploit artificial neural networks to construct trial wavefunctions, where the connection weights in the network play the role of variational parameters \cite{CarleoScience2017,GlasserPRX2018}. Neural network approaches have already been succesffuly applied to a wide number (see e.g. \cite{Nieuwenburg2017,Schindler2017,Choo2018,Czischek2018,DirectSampling2019}) of close Hamiltonian systems. However, they have not yet been generalized to the important quantum manybody problem of open systems. 

In this Letter, we present a theoretical approach based on a variational neural network ansatz in order to determine the steady state of the master equation of open quantum lattice systems. We construct the ansatz for the mixed density matrix starting from a  Restricted Boltzman Machine ansatz for a pure many-body wavefunction in an extended Hilbert space. 
We determine the optimal variational parameters by minimizing a cost function which involves the Liouvillian superoperator associated to the master equation for the density matrix. As a first application, we have considered the dissipative tranverse field quantum Ising model. We present a proof-of-principle demonstration by  benchmarking the neural network calculations of the steady state against numerically exact simulations performed by quantum trajectories in the full Hilbert space \cite{Daley2014}. Our minimization of the cost function is performed by Markov chain Monte Carlo sampling of the gradient and is thus scalable to a large number of lattice sites. Perspectives of the present approach are discussed in the conclusions.

The general task that we wish to solve is the determination of the steady state of an open quantum system described by the Lindblad master equation \cite{Breuer2007}
for the system reduced density matrix $\orho$, which reads (setting $\hbar=1$):
\begin{equation}
	\label{eq:me}
	\dot{\orho}=\liouv\orho = -i\comm{\oH}{\orho}+\sum_j\frac{\gamma_j}{\new{2}}\left[2\hat{L}_j\orho\hat{L}^\dagger_j - \{\hat{L}_j^\dagger\hat{L}_j, \orho \} \right],
\end{equation}
where $\liouv$ is the so-called Liouvillian superoperator depending on the system Hamiltonian operator $\oH$. The coupling to the environment is represented by interaction channels with the reservoir characterized by dissipation rates $\gamma_j$ and jump operators $\hat{L}_j$ acting on the system. 
\new{Here we will focus on situations where the steady state ($\partial_t \orho_{ss} =0$) is unique. In this case, the  steady-state density matrix  can be obtained as $\orho_{ss}=\lim_{t\rightarrow\infty} \orho(t)$ regardless of the initial condition. 
Although it is possible to engineer peculiar Liouvillians with more than one steady state \cite{AlbertPRA14}, typical physical systems with a finite Hilbert space dimension have a unique steady-state \cite{Spohn1977,PhysRevA.98.042118,Nigro_2019Nigro_2019}. }

\new{For the many-body problem an analytical expression for $\orho_{ss}$ can be found in very few cases \cite{Prosen2014PRL,Prosen2011PRL}.}
\new{In general,} because of the exponential growth of the Hilbert space with the number of lattice sites, describing the full density matrix requires exponentially many complex numbers, which in practice can be done exactly only for a small number of sites. 
If one wants to attack the problem within a variational framework, the density matrix can be represented by an ansatz  $\hat\rho_{\bm v}$ depending on a set of variational parameters ${\bm v}$. If $\{ \ket{{\bm \sigma}} = \ket{\sigma_1,\sigma_2,...,\sigma_N} \}$ denotes a basis of states for the system Hilbert space, the density matrix can be expressed in the  form
\begin{equation}
	\label{eq:rho-components}
	\hat\rho({\bm v}) = \sum_{\bm{\sigma}, \bm{\sigma'}} \rho_{\bm{v}} ({\bm \sigma}, \bm{\sigma'}) \ket{\bm \sigma}\bra{\bm \sigma'}.
\end{equation}

\begin{figure}
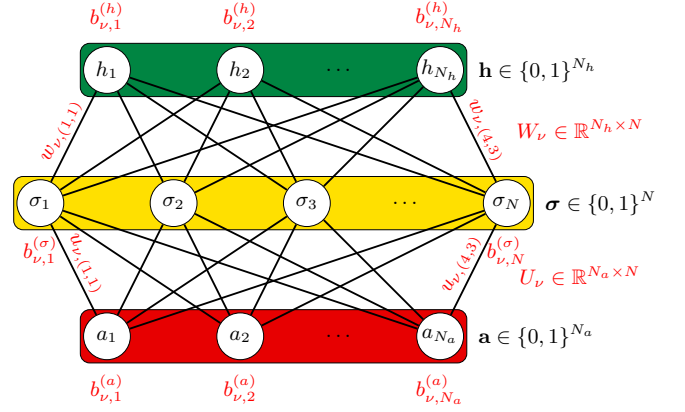

\begin{center}
\includestandalone[width=\columnwidth]{NDM}
\end{center}
\caption{Graph representation of the artificial neural network used for the density matrix ansatz. The vector ${\bm \sigma} = (\sigma_1,\sigma_2,...,\sigma_N)$ contains the variables of the physical system (visible layer). The vector ${\bm a} = (a_1,a_2,...,\sigma_{N_a})$ describes the ancillary degrees of freedom of the extended Hilbert space (ancilla layer), where a purified density matrix is considered (see Eq. (\ref{eq:purification-ansatz})). The vector ${\bm h} = (h_1,h_2,...,h_{N_h})$ contains variables of auxiliary nodes (hidden layer).  The network parameters are ${\bm v}_{\nu} = ({\bf{b}}^{(\sigma)}_{\nu},{\bm b}^{(a)}_{\nu},{{\bm b}}^{(h)}_{\nu},{{\bm W}}_{\nu},{{\bm U}}_{\nu})$. One network is used for the representation of the amplitude ($\nu = \vv_A$) of the purified wavefunction in Eq. (\ref{eq:purification-ansatz}), while another independent network with the same topology is used to represent the phase ($\nu = \vv_\theta$).  	\label{fig:ndm}}
\end{figure}

In order to construct our neural network ansatz for the density matrix, we consider an extended Hilbert space $\mathcal{H}=\mathcal{H}_S\otimes\mathcal{H}_A$ where $\mathcal{H}_{S,A}$ represents respectively the system and ancillary Hilbert spaces. Such extended space is spanned by the basis set $\{ \ket{{\bm \sigma},{\bm a} } \}$ where ${\bm a} = ({a_1,a_2,...,a_{N_a}})$ labels the ancillary degrees of freedom. We start by considering a pure state in the extended Hilbert space, represented by the wavefunction  $\psi_{\bm v}({\bm \sigma}, {\bm a})$. In this framework the reduced density matrix of the system $S$ is obtained by tracing out the ancillary degrees freedom \cite{Torlai2018}, namely
\begin{equation}
	\label{eq:purification-ansatz}
	\rho_{\bm v}(\bm{\sigma},{\bm \sigma'}) = \sum_{\bm a} \psi_{\bm v}({\bm \sigma}, {\bm a}) \psi_{\bm v}^\star({\bm \sigma'}, {\bm a}).
\end{equation}
The next step is to represent $\psi_{\bm v}({\bm \sigma}, {\bm a})$ via a neural network ansatz.
This purified procedure automatically ensures that $\hat\rho_{\bm v}$   is Hermitian and positive semi-definite, as required for a density matrix.
In a recent paper, Torlai and Melko \cite{Torlai2018} proposed to describe purified wavefunctions as 
\begin{equation}
\label{eq:wave}
\psi_{\bm v}({\bm \sigma},{\bm a}) = \sqrt{{\mathcal P}_{{\bm v}_A}({\bm \sigma},{\bm a})}\exp[-1/2\log({\mathcal P}_{{\bm v}_{\theta}}({\bm \sigma},{\bm a}))].
\end{equation} Both the amplitude ${\mathcal P}_{{\bm v}_A}({\bm \sigma},{\bm a})$ and phase-related function ${\mathcal P}_{{\bm v}_{\theta}}({\bm \sigma},{\bm a})$ of the purified wavefunction are given by the Boltzmann-like expression ${\mathcal P}_\nu({\bm \sigma}, {\bm a}) = \sum_{\bm h} \exp[-E_\nu({\bm \sigma}, {\bm a}, {\bm h})]$ (with $\nu \in \{\vv_A,\vv_\theta \}$), where the associated dimensionless energy reads
\begin{equation}
	\label{eq:energy-ndm}
	E_\nu({\bm \sigma},{\bm a}, {\bm h}) = {\bm \sigma}\cdot {\bf{b}}^{(\sigma)}_{\nu} + {\bm a}\cdot {\bm b}^{(a)}_{\nu} + {\bm h}\cdot {{\bm b}}^{(h)}_{\nu} + {\bm \sigma}^{T} {{\bm W}}_{\nu}  {\bm h} + {\bm \sigma}^{T} {{\bm U}}_\nu {\bm a}.
\end{equation}
Note that the ansatz parameters are ${\bm v} = ({\bm v}_A,{\bm v}_{\theta})$ where ${\bm v}_{\nu} = ({\bf{b}}^{(\sigma)}_{\nu},{\bm b}^{(a)}_{\nu},{{\bm b}}^{(h)}_{\nu},{{\bm W}}_{\nu},{{\bm U}}_{\nu})$. The rectangular matrix ${{\bm W}}_{\nu}$ weighs the connections between the system variables (visible layer) to the auxiliary variables (hidden layer), while the weight matrix ${{\bm U}}_{\nu}$ quantifies the connection between the system variables and the ancillary ones (ancillary layer). 
Such neural network ansatz is represented by a tri-partite Restricted Boltzmann Machine depicted in  \cref{fig:ndm}. 
In other words, there are two independent artificial neural networks, one for the amplitude ($\nu = A$)  and one for the phase ($\nu = \theta$).
By substituting those formulas into \cref{eq:purification-ansatz} and carrying out the sum over the ancillary degrees of freedom one obtains a closed formula for the entries of the density matrix:
\begin{equation}
	\rho_{{\bm v}}({{\bm \sigma}, {\bm \sigma'}})= \exp[\Gamma^-_{\vv}(\bm{ \sigma}, \bm{\sigma'}) + \Gamma^+_{\vv}({{\bm \sigma}, {\bm \sigma'}}) + \Pi_{\vv}({{\bm \sigma}, {\bm \sigma'}})]
\end{equation}

where the expression of $\Gamma^{+/-}$ and $\Pi$ can be found in the Supplemental Material \footnote{For more details, see Supplemental Material at the end of this PDF document}.
\new{The representation power \cite{LeRoux2008NeurComp,Younes1996AML,Montufar2011NeurComp} of this ansatz can be systematically improved by increasing the density of the hidden ($\alpha=N_h/N$) and ancillary layer ($\beta=N_a/N$).}
It is worth pointing out that this scheme is not specific to this network topology, but relies only on the general fact that if two visible layers are connected by a shallow ancillary layer, the ancilla can be traced out analitically and an efficient neural-network description of the density matrix can be obtained. 

Having defined a variational ansatz $\orho(\bm v)$, we now wish to define a variational principle to determine the optimal parameters.
In particular, we have to recast the search for the steady state into a minimization problem for a real, positive cost function $\mathcal{C}({\bm v}) $  which has a global minimum when the master equation ${\mathcal L} \orho_{{\bm v}} = 0$ is satisfied \cite{Weimer2015}.
Moreover,  in order to be able to deal with large Hilbert spaces, we need a quantity which can be sampled and computed efficiently.
These requirements are met by the following cost function expressed in terms of the 2-norm of the time derivative of the density-matrix:
\begin{equation}
	\label{eq:l2-cost-function}
	\mathcal{C}({\bm v}) = \frac{\norm{d\orho_{\bm v}/dt}^2_2}{\norm{\orho_{\bm v}}^2_2} = \frac{\Tr[\orho^\dagger_{\bm v}\liouvd\liouv\orho_{\bm v}]}{\Tr[\orho^\dagger_{\bm v}\orho_{\bm v}]},
\end{equation}
as  (i) $\mathcal{C}({\bm v}_{ss}) = 0 \iff \orho({\bm v}_{ss}) =\orho_{ss}$ and (ii) $\mathcal{C}({\bm v})\geq 0$. 

It is useful to rewrite \cref{eq:l2-cost-function} as a sum over the whole space of bounded operators on the Hilbert space: 
\begin{equation}
	\mathcal{C}({\vv}) = \sum_{\vsigma, \vtsigma}  p_{\vv}(\vsigma, \vtsigma)  \new{|}\mathcal{C}^\text{Loc}(\vv,\vsigma, \vtsigma)\new{|^2},
\end{equation}
where $p_{\vv}(\vsigma, \vtsigma) =\abs{\rho_{\vv}(\vsigma, \vtsigma)}^2/Z$ corresponds to a probability distribution as $Z=\sum_{\vsigma, \vtsigma} \abs{\rho_{\vv}(\vsigma, \vtsigma)}^2 $ \footnote{Being able to rewrite the cost function (\ref{eq:l2-cost-function}) into this form allows us to sample it efficiently. 
This is the main advantage with respect to using the trace norm of $\liouv\orho$ \cite{Weimer2015} as a cost function, which requires computing the singular-value-decomposition of $\liouv\rho$ at each iteration.}. 
The local contribution reads \footnote{\new{
This is not the only possible expression that allows to sample the cost function $\mathcal{C}(\bm{v})$.
However, this choice of the local cost function $\mathcal{C}^{\rm Loc}$ is particularly convenient since it respects the zero-variance property, it is more numerically stable and cheaper to compute. For further details see Sec. 3 of the Supplemental Material.}} :
\begin{equation}
\label{eq:cost-fun-local}
\mathcal{C}^\text{Loc} (\vv, \vsigma, \vtsigma)= \sum_{\vsigma',\vtsigma'} \new{\mathcal{L}(\vsigma, \vtsigma; \vsigma', \vtsigma') \frac{\rho_{\vv}(\vsigma', \vtsigma')}{\rho_{\vv}(\vsigma, \vtsigma)}} .
\end{equation}
In order to find the global minimum of the cost function (\ref{eq:l2-cost-function}) by means of gradient-based iterative schemes we need to compute its gradient 
\new{\begin{multline}
	\nabla_{\vv}\ \mathcal{C}({\vv}) = \sum_{\vsigma, \vtsigma}  p_{\vv}(\vsigma, \vtsigma) \mathcal{C}^\text{Loc}(\vv,\vsigma, \vtsigma)^\star \Big[\sum_{\sigma',\tilde{\sigma}'}\mathcal{L}(\vsigma, \vtsigma; \vsigma', \vtsigma')\cr 
	\frac{\rho_{\vv}(\vsigma', \vtsigma')}{\rho_{\vv}(\vsigma, \vtsigma)} \mathcal{O}_{\vv}(\vsigma',\vtsigma')\Big]  - \new{\mathcal{C}(\vv)}\mathcal{O}_{\vv},
\end{multline}}
where we have defined the log-derivatives of the density matrix $\mathcal{O}_{\vv}=\sum_{\vsigma, \vtsigma} \mathcal{O}_{\vv}(\vsigma, \vtsigma)$ and $\mathcal{O}_{\vv}(\vsigma, \vtsigma) = \nabla_{\vv} \log{\rho_{\vv}(\vsigma, \vtsigma)}$, which can be efficiently computed for the considered neural network. 

\begin{figure}
\begin{center}
\includegraphics[width=0.8\columnwidth]{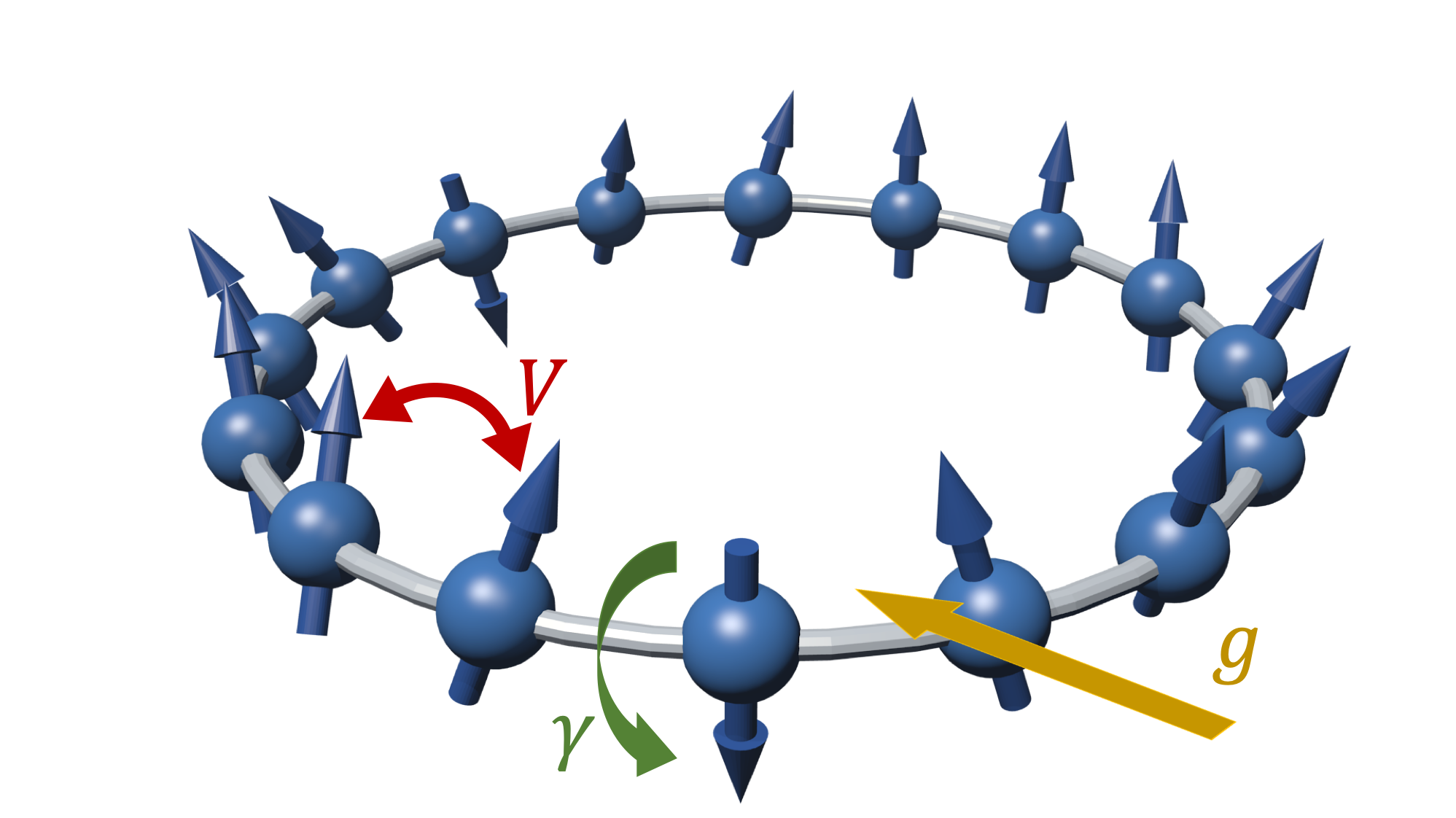}
\includegraphics[width=0.9\columnwidth]{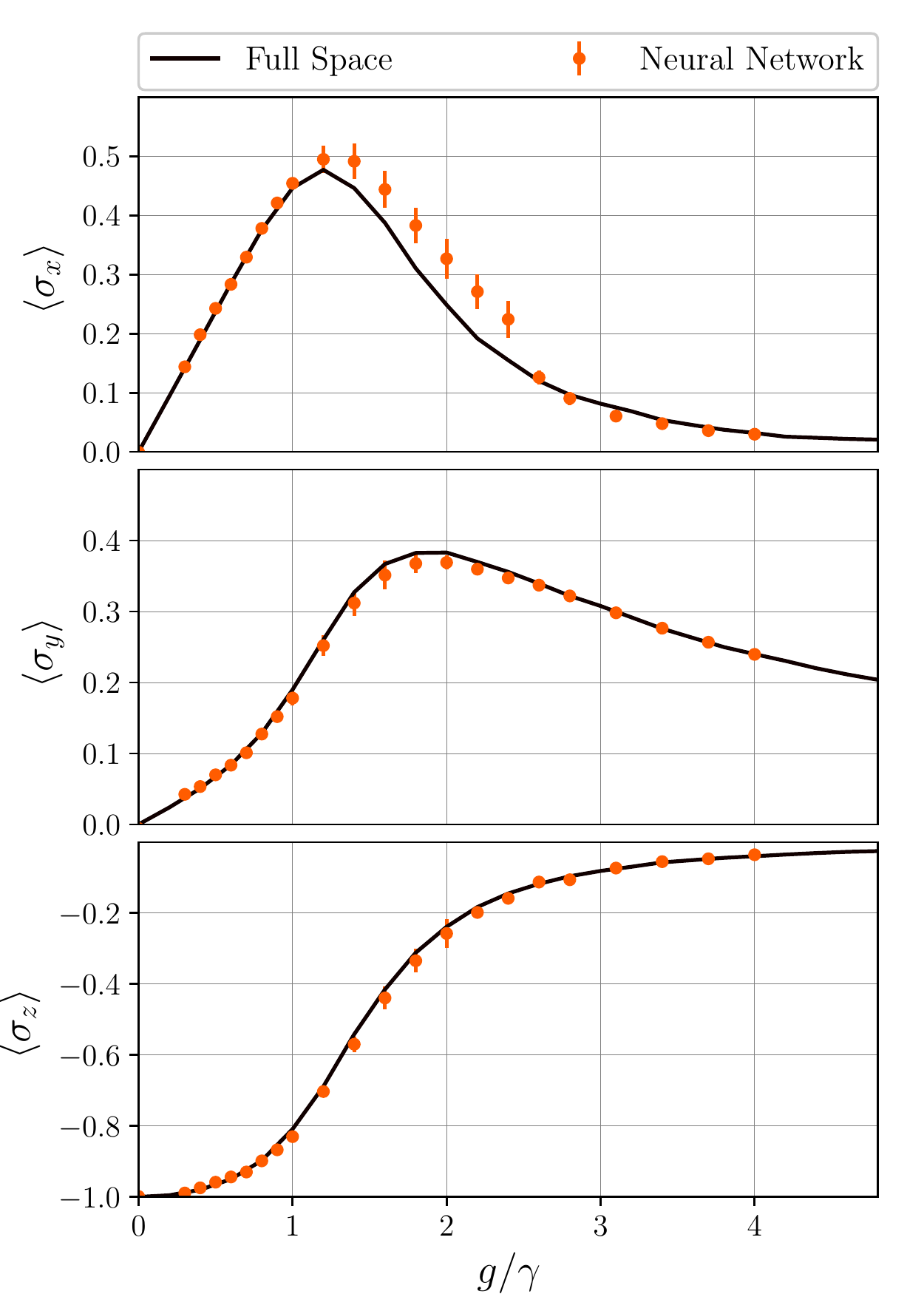}
\end{center}
\caption{Top panel: sketch of the considered physical system described by the dissipative quantum transverse Ising 1D model with periodic boundary conditions. The quantity  $g$ denotes an applied magnetic field, $V$ the spin-spin coupling and  $\gamma$ the spin flip rate. 
Bottom panel: the prediction of the neural-network variational calculations (circles) are compared to the results obtained by quantum trajectory simulations of the master equation by considering the whole Hilbert space (solid lines). The top, middle and bottom panels depict the expectation values of the three components of the averaged magnetization as a function of the applied magnetic field $g$ (in units of $\gamma$). Model parameters: $V/\gamma = 2$ (spin-spin coupling), $N = 16$ (number of lattice sites). Neural-network parameters: \new{$\alpha=\beta=1$ for $g \leq \gamma$ and $g \geq 2.5 \gamma$ while $\alpha=1$ and $\beta=4$ for the remaining points.}
The parameters  required for the convergence of the Monte Carlo calculations depend on the value of $g/\gamma$, with the intermediate region being the most demanding.  The maximum number of accepted Monte Carlo samples is \new{$8640$} and the maximum number of steps for the stochastic gradient descent is $10^4$. For points outside the intermediate region \new{$3000$} accepted Monte Carlo samples  and $10^3$ iteration steps have been performed.
	\label{fig:benchmark}}
\end{figure}

The computational complexity of evaluating $\nabla\mathcal{C}({\vv})$ exactly grows exponentially with the size of the system. 
This cost can be considerably reduced if one only uses an estimate of $\nabla\mathcal{C}({\vv})$ obtained by sampling the values $(\vsigma, \vtsigma)$ according to the probability $p_{\vv}(\vsigma, \vtsigma)$.
Because the normalisation factor $Z$ is not fixed, we cannot sample the distribution directly and have to resort to a Markov Chain Monte Carlo \cite{Becca2017} with Metropolis update rules \footnote{We point out that a promising direction of research would be to devise particular trial wave functions where $Z$ is fixed or cheaper to compute, so that a direct sampling of the distribution $p_{\sigma, \tsigma}$ without a Markov Chain would lead to easier convergence properties.
Indeed, it has been recently shown \cite{DirectSampling2019} that a direct sampling is possible in some types of networks.}. 
At every sampling step, we propose to update the configuration $(\vsigma, \vtsigma) \rightarrow (\vsigma', \vtsigma')$ by switching a random number of spins  and accept the new configuration with probability $\min(\exp[p_{\vv}(\vsigma, \vtsigma)/p_{\vv}(\vsigma', \vtsigma')], 1)$. 

Finally, in order to find the global minimum of the cost function we employ a standard Stochastic Gradient Descent  algorithm \cite{StochasticGradientDescent}. 
In order to improve the performance of the Stochastic Gradient Descent (i.e. to reduce the number of iterations needed to converge to the global minima of the cost function) we update the variational parameters according to to the metric of the space of density matrices exploiting the Stochastic Reconfiguration Approach \cite{BeccaBookChOptimization2017}.
During the optimization procedure we sample the physical observables of interest through another Markov chain as
\begin{equation}
\expval*{\hat\Theta} = \frac{\Tr[\orho \ \hat\Theta]}{\Tr[\orho]} = \sum_{\vsigma} p^{\rm obs}_{\vv}(\vsigma) \sum_{\vtsigma}\frac{\rho_{\vv}(\vsigma,\vtsigma)\Theta(\vtsigma,\vsigma)}{\rho_{\vv}(\vsigma,\vsigma)},
\end{equation}
where $p^{\rm obs}_{\vv}(\vsigma)=\rho_{\vv}(\vsigma,\vsigma)/\Tr[\orho]$.

In order to benchmark our neural-network approach for open quantum systems, we consider here the dissipative quantum transverse Ising model, whose Hamiltonian is
\begin{equation}
	\label{eq:quantum-ising}
	H=\frac{V}{4}\sum_{\langle j,l\rangle} \hat\sigma^z_j\hat\sigma_l^z + \frac{g}{2}\sum_j \hat\sigma_j^x ,
\end{equation}
being $ \hat\sigma_j^{\alpha}$ the Pauli matrices ($\alpha \in \{x,y,z \}$) acting on the $j$-th site.
The first term represents the nearest-neighbor spin-spin interaction depending only on the $z$-components, being $V$ the coupling strength. The second term accounts for a local and uniform magnetic field along the transverse direction $x$. 
We consider local dissipative spin-flip processes described  by the site-dependent jump operator $\hat{L}_j^{(z)} = \hat\sigma_j^-=\frac{1}{2}(\hat\sigma^x_j - i\hat\sigma^y_j)$, which fully determine the master equation in Eq. (\ref{eq:me}).

\begin{figure}
\begin{center}
\includegraphics[width=0.9\columnwidth]{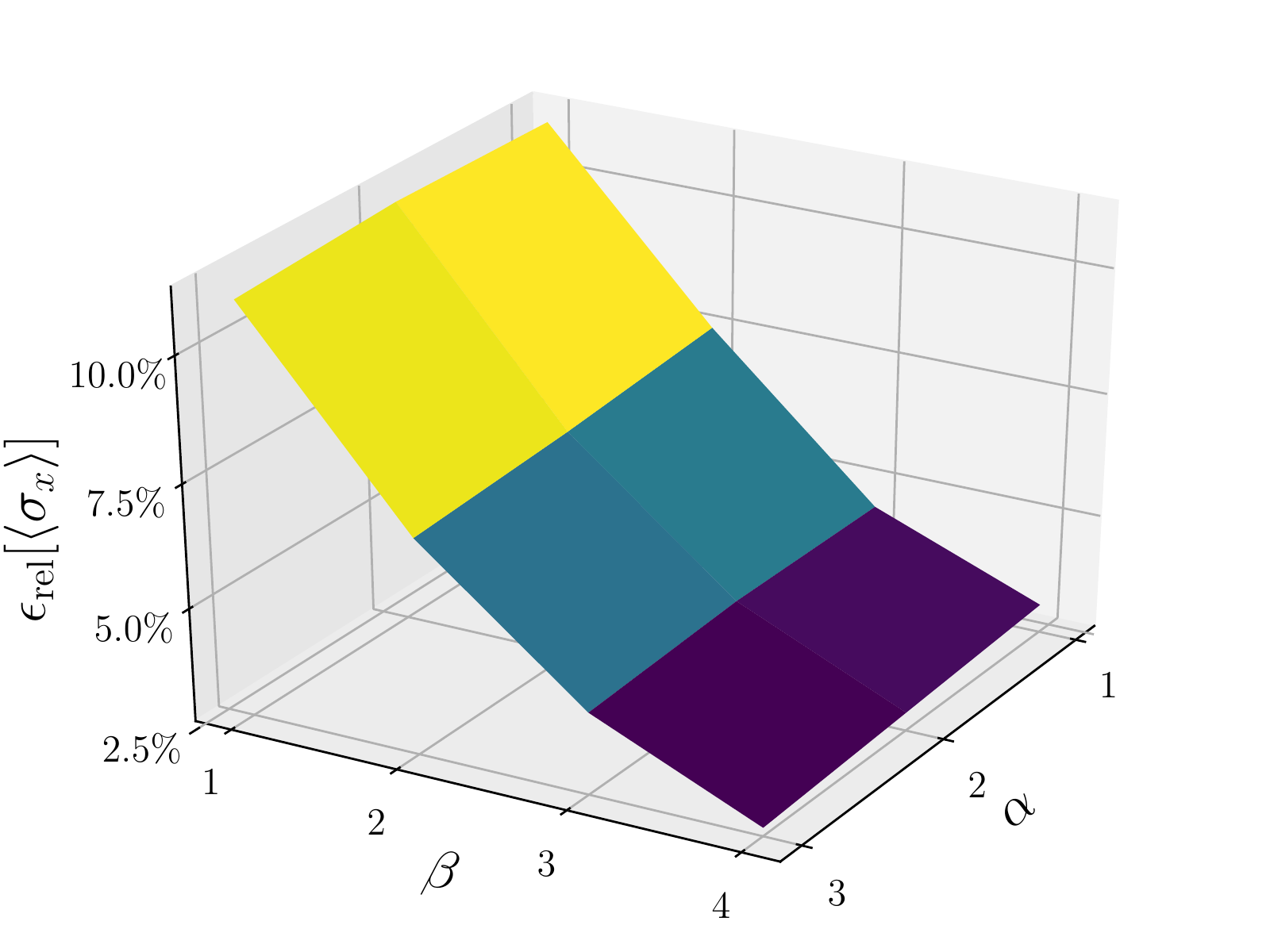}
\end{center}
\caption{\new{Relative error with respect to the exact result for the observable $\expval{\sigma_x}$ as a function of $\alpha$ and $\beta$. Parameters are set as in Fig.\ref{fig:benchmark} but for a fixed value $g/\gamma = 1.2$.}	\label{fig:error}}
\end{figure}

\begin{figure}
\begin{center}
\includegraphics[width=0.9\columnwidth]{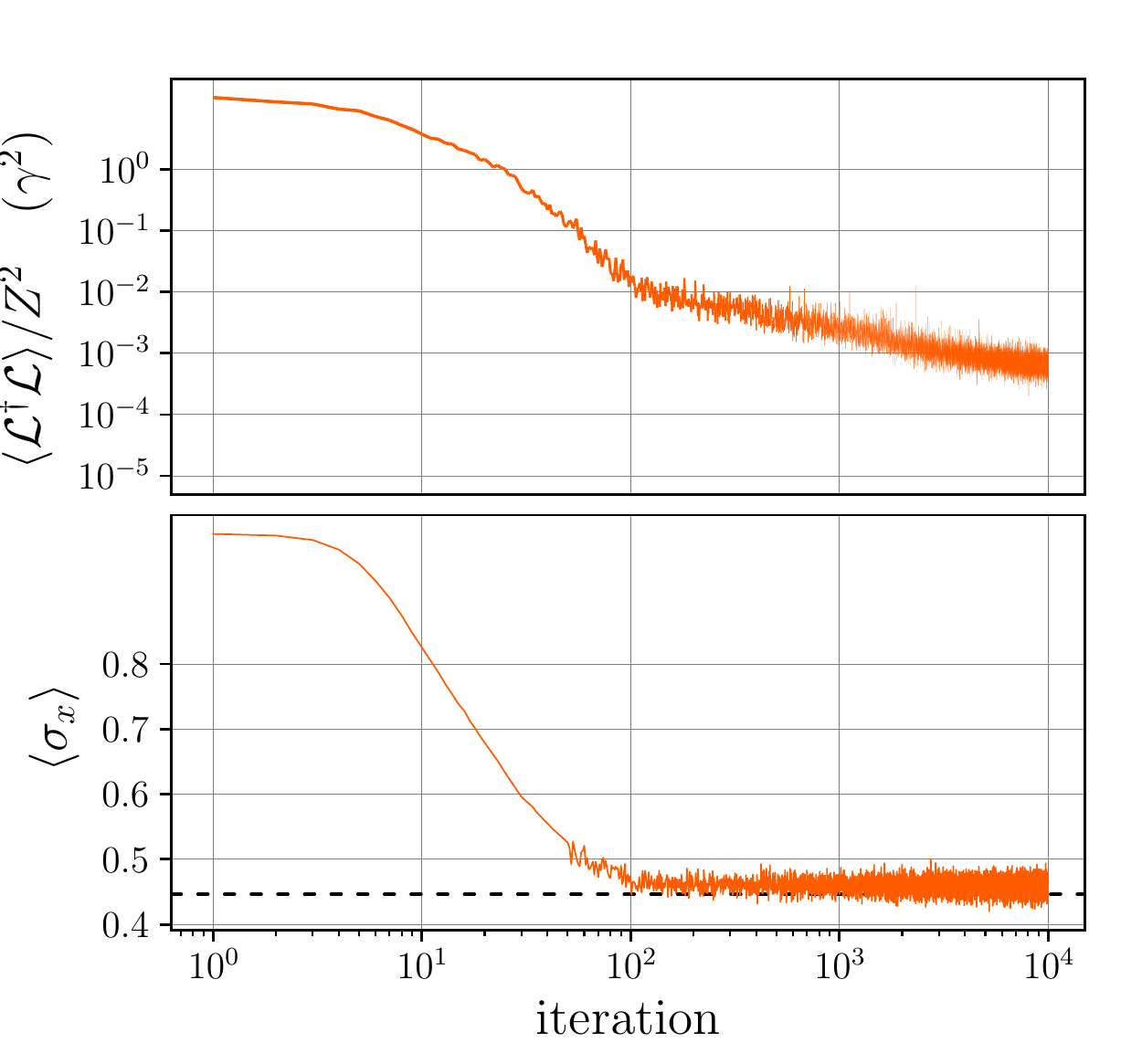}
\end{center}
\caption{Same parameters as in Fig. \ref{fig:benchmark} but for a fixed value $g/\gamma = 1$.
Top panel: the cost function is shown as a function of the iteration steps. Bottom panel: the corresponding  evolution of 
the $x$-component of the average magnetization during the stochastic minimization is shown.
	\label{fig:convergence}}
\end{figure}

Numerical results for steady-state observables of the dissipative quantum transverse Ising model on a 1D periodic chain are reported in Fig. \ref{fig:benchmark}.
In particular, we report the spatial components of the averaged magnetization as a function of the magnetic field $g$ (in units of the dissipation rate $\gamma$)
for $V/\gamma = 2$.
For $N = 16$ lattice sites the predictions of the neural-network variational method (circles) are compared to the results obtained with
a brute-force exact integration of the master equation in the whole Hilbert space, showing a 
\new{good agreement over all the parameter range.
For $g \lesssim \gamma$ and $g \gtrsim 2.5 \gamma$ a remarkable precision is reached for all the local observables with a low density of the hidden and ancillary layer $\alpha=\beta=1$ and $\mathcal{O}(10^2)$ minimization steps.}
\new{For $1\lesssim g / \gamma\lesssim2.5$ an higher number of variational parameters is required.
In particular, as shown in Fig. \ref{fig:error}, a systematic improvement of the relative error $\epsilon_{\rm rel}\big[\expval{\sigma_x}\big]$ with respect to the exact solution can be obtained by increasing $\beta$.}
Interestingly, for $1\lesssim g / \gamma\lesssim2.5$, we note that the gradient-descent procedure requires more iterations. This region corresponds to the range of $g/\gamma$ where the smallest nonzero eigenvalue of $\liouvd\liouv$ decreases significantly \cite{Jin2018_ising}.   
\new{In this range the steady-state density matrix also displays nontrivial correlations and non-thermal mixness properties \cite{Jin2018_ising}.
Remarkably, the fidelity of the reconstructed local density matrix with respect to the exact one is alway larger then $0.998$ for all the values of $g/\gamma$ considered.}
Finally, as an example of convergence, the top panel of Fig. \ref{fig:convergence} depicts a typical evolution of the cost function in the iterative minimization procedure for a fixed set of parameters ($g/\gamma=1$), showing a good convergence towards the global minimum. In the bottom panel of  Fig. \ref{fig:convergence}, the convergence of the $x$-component of the averaged magnetization is also reported.

In conclusion, we have presented a general variational approach for the steady-state density matrix of open quantum manybody systems based on an artificial neural network scheme.
Our method is scalable since the cost function associated to the Liouvillian of the master equation can be calculated via Monte Carlo sampling. We have demonstrated a proof-of-principle of the theoretical scheme by a successful benchmarking to brute-force finite-size simulations in the full Hilbert space for arrays of spins described by the dissipative quantum transverse Ising model.  
We would like to point out that the present approach does not depend on the specific network topology. Indeed, the variational procedure presented in this Letter is general and can be applied to many other neural-networks or physically-inspired variational ans\"{a}tze. There are many future developments at the horizon, including the study of dynamical properties, the use of deep neural networks and/or alternative cost functions, comparison with other existing techniques as well as the study of disordered systems without translational invariance. The neural network approach has the potential to pave the way to the theoretical study of a wide spectrum of open quantum manybody systems. 

\acknowledgements
We thank G. Carleo, V. Savona and G. Orso for fruitful discussions.
Numerical code for this paper has been written in \textit{Julia} \cite{Julia2017}. 
Full space simulations have been made with \textit{QuantumOptics.jl} \cite{quantumopticsjl} and with \textit{QuTiP} \cite{qutip2012,qutip2013}.
We acknowledge support from ERC (via Consolidator Grant CORPHO No. 616233). This work was granted access to the HPC resources of TGCC under the allocation 2018-A0050510601 attributed by GENCI (Grand Equipement National de Calcul Intensif).

{\it Note: while completing this work, we became aware of related independent theoretical works that have been carried on in parallel \cite{Hartmann-2019arXiv190205131H,giappi19arxiv,Nagy19arxiv}}.

\bibliographystyle{apsrev4-1}
\bibliography{biblio}

\clearpage

\onecolumngrid


\section*{Supplementary material}

# Supplemental online material for the paper
# *Variational neural network ansatz for steady states in open quantum systems*

Filippo Vicentini,[1] Alberto Biella,[1] Nicolas Regnault,[2] and Cristiano Ciuti[1]

[1]*Laboratoire Matériaux et Phénomènes Quantiques,*
*Université Paris Diderot, CNRS-UMR7162, 75013 Paris, France*

[2]*Laboratoire de Physique de l'Ecole Normale Supérieure,*
*ENS, Université PSL, CNRS, Sorbonne Université,*
*Université Paris-Diderot, Sorbonne Paris Cité, Paris, France*

(Dated: May 27, 2019)

## I. NEURAL NETWORK ANSATZ

We provide here the detailed expressions of the quantities $\Gamma_{\boldsymbol{v}}^{+}(\boldsymbol{\sigma},\tilde{\boldsymbol{\sigma}})$, $\Gamma_{\boldsymbol{v}}^{-}(\boldsymbol{\sigma},\tilde{\boldsymbol{\sigma}})$ and $\Pi_{\boldsymbol{v}}(\boldsymbol{\sigma},\tilde{\boldsymbol{\sigma}})$ entering the neural network ansatz $\rho_{\boldsymbol{v}}(\boldsymbol{\sigma},\tilde{\boldsymbol{\sigma}}) = \exp[\Gamma_{\boldsymbol{v}}^{+}(\boldsymbol{\sigma},\tilde{\boldsymbol{\sigma}}) + \Gamma_{\boldsymbol{v}}^{-}(\boldsymbol{\sigma},\tilde{\boldsymbol{\sigma}}) + \Pi_{\boldsymbol{v}}(\boldsymbol{\sigma},\tilde{\boldsymbol{\sigma}})]$:

$$\Gamma_{\boldsymbol{v}}^{+}(\boldsymbol{\sigma},\tilde{\boldsymbol{\sigma}}) = \frac{1}{2}\left[\boldsymbol{b}_{\boldsymbol{A}}^{(v)} \cdot (\boldsymbol{\sigma}+\tilde{\boldsymbol{\sigma}}) + \sum_j \log\left[\mathcal{G}(\theta_A^{[j]}(\boldsymbol{\sigma}))\right] + \sum_j \log\left[\mathcal{G}(\theta_A^{[j]}(\tilde{\boldsymbol{\sigma}}))\right]\right], \quad (1)$$

$$\Gamma_{\boldsymbol{v}}^{-}(\boldsymbol{\sigma},\tilde{\boldsymbol{\sigma}}) = \frac{i}{2}\left\{\boldsymbol{b}_{\boldsymbol{\theta}}^{(v)} \cdot (\boldsymbol{\sigma}-\tilde{\boldsymbol{\sigma}}) + \sum_j \log\left[\mathcal{G}(\theta_\theta^{[j]}(\boldsymbol{\sigma}))\right] - \sum_j \log\left[\mathcal{G}(\theta_\theta^{[j]}(\tilde{\boldsymbol{\sigma}}))\right]\right\}, \quad (2)$$

$$\Pi_{\boldsymbol{v}}(\boldsymbol{\sigma},\tilde{\boldsymbol{\sigma}}) = \sum_k \log\left\{\mathcal{G}\left[1/2(\boldsymbol{\sigma}+\tilde{\boldsymbol{\sigma}})^T \boldsymbol{U}_{\boldsymbol{A}}^{[k]} + i/2(\boldsymbol{\sigma}-\tilde{\boldsymbol{\sigma}})^T \boldsymbol{U}_{\boldsymbol{\theta}}^{[k]} + b_{A,k}^{(a)}\right]\right\}, \quad (3)$$

where $U_\nu^{[k]}$ is the k-th column of the $U_\nu$ matrix, and $\theta_\nu^{[j]}$ is the j-th component of the $\theta_\nu$ vector

$$\boldsymbol{\theta_\nu}(\boldsymbol{\sigma}) = \boldsymbol{b}_\nu^{(h)} + \boldsymbol{\sigma}^T W_\nu. \quad (4)$$

$\mathcal{G}(x)$ is a non-linear activation function dependent on the values that hidden spins can take. In the manuscript we consider binary spins ($h_i = \{0,1\}$), so $\mathcal{G}(x) = 1 + e^{-x}$.

The total number of real variational parameters of the neural network ansatz (the length of the vector $\boldsymbol{v}$) is given by the following formula:

$$N_{\text{par}} = dim(\boldsymbol{v}) = 2(N + \alpha N + \beta N + \alpha N^2 + \beta N^2), \quad (5)$$

where $\alpha = N_h/N$ and $\beta = N_a/N$ are the hidden and ancillary layer density. For $N \gg 1$ and assuming $\beta = 1$ as in the main text, we get $N_{\text{par}} \sim \mathcal{O}\left((\alpha+1)N^2\right)$. The number of parameters can be compared to a Matrix Product Operator representation of the density matrix where $N_{\text{par}}^{\text{MPO}} \sim \mathcal{O}(N\chi^2)$ where $\chi$ is the bond-link dimension [1], playing a role similar to $\alpha$. For comparison, an exact description in the full Hilbert space requires $\mathcal{O}(2^{2N})$ parameters.

## II. LOG-DERIVATIVES

The logarithmic derivatives of the variational density matrix $\rho_{\boldsymbol{v}}(\sigma,\tilde{\boldsymbol{\sigma}})$ appearing in Eq. (10) of the main text are computed with respect to the various components of $\boldsymbol{v}$ as follows:

$$\frac{\partial \log[\rho_{\boldsymbol{v}}(\boldsymbol{\sigma}, \tilde{\boldsymbol{\sigma}})]}{\partial \mathbf{b}_A^{(v)}} = \frac{1}{2}(\boldsymbol{\sigma} + \tilde{\boldsymbol{\sigma}}), \tag{6}$$

$$\frac{\partial \log[\rho_{\boldsymbol{v}}(\boldsymbol{\sigma}, \tilde{\boldsymbol{\sigma}})]}{\partial \mathbf{b}_\theta^{(v)}} = \frac{i}{2}(\boldsymbol{\sigma} - \tilde{\boldsymbol{\sigma}}), \tag{7}$$

$$\frac{\partial \log[\rho_{\boldsymbol{v}}(\boldsymbol{\sigma}, \tilde{\boldsymbol{\sigma}})]}{\partial b_A^{(h)[j]}} = \frac{1}{2}\left\{\frac{\mathcal{G}'}{\mathcal{G}}[\theta_A^{[j]}(\boldsymbol{\sigma})] + \frac{\mathcal{G}'}{\mathcal{G}}[\theta_A^{[j]}(\tilde{\boldsymbol{\sigma}})]\right\}, \tag{8}$$

$$\frac{\partial \log[\rho_{\boldsymbol{v}}(\boldsymbol{\sigma}, \tilde{\boldsymbol{\sigma}})]}{\partial b_\theta^{(h)[j]}} = \frac{i}{2}\left\{\frac{\mathcal{G}'}{\mathcal{G}}[\theta_\theta^{[j]}(\boldsymbol{\sigma})] - \frac{\mathcal{G}'}{\mathcal{G}}[\theta_\theta^{[j]}(\tilde{\boldsymbol{\sigma}})]\right\}, \tag{9}$$

$$\frac{\partial \log[\rho_{\boldsymbol{v}}(\boldsymbol{\sigma}, \tilde{\boldsymbol{\sigma}})]}{\partial b_A^{(a)[j]}} = \frac{\mathcal{G}'}{\mathcal{G}}\left[\frac{1}{2}(\boldsymbol{\sigma} + \tilde{\boldsymbol{\sigma}}) \cdot \mathbf{U}_A^{[j]} + \frac{i}{2}(\boldsymbol{\sigma} - \tilde{\boldsymbol{\sigma}}) \cdot \mathbf{U}_\theta^{[j]} + b_{A,j}^{(a)}\right], \tag{10}$$

$$\frac{\partial \log[\rho_{\boldsymbol{v}}(\boldsymbol{\sigma}, \tilde{\boldsymbol{\sigma}})]}{\partial \mathbf{W}_A^{[j]}} = \frac{1}{2}\left\{\frac{\mathcal{G}'}{\mathcal{G}}[\theta_A^{[j]}(\boldsymbol{\sigma})]\,\boldsymbol{\sigma} + \frac{\mathcal{G}'}{\mathcal{G}}[\theta_A^{[j]}(\tilde{\boldsymbol{\sigma}})]\,\tilde{\boldsymbol{\sigma}}\right\}, \tag{11}$$

$$\frac{\partial \log[\rho_{\boldsymbol{v}}(\boldsymbol{\sigma}, \tilde{\boldsymbol{\sigma}})]}{\partial \mathbf{W}_\theta^{[j]}} = \frac{i}{2}\left\{\frac{\mathcal{G}'}{\mathcal{G}}[\theta_\theta^{[j]}(\boldsymbol{\sigma})]\,\boldsymbol{\sigma} - \frac{\mathcal{G}'}{\mathcal{G}}[\theta_\theta^{[j]}(\tilde{\boldsymbol{\sigma}})]\,\tilde{\boldsymbol{\sigma}}\right\}, \tag{12}$$

$$\frac{\partial \log[\rho_{\boldsymbol{v}}(\boldsymbol{\sigma}, \tilde{\boldsymbol{\sigma}})]}{\partial \mathbf{U}_A^{[j]}} = \frac{1}{2}\,(\boldsymbol{\sigma} + \tilde{\boldsymbol{\sigma}})\,\frac{\mathcal{G}'}{\mathcal{G}}\left[\frac{1}{2}(\boldsymbol{\sigma} + \tilde{\boldsymbol{\sigma}}) \cdot \mathbf{U}_A^{[j]} + \frac{i}{2}(\boldsymbol{\sigma} - \tilde{\boldsymbol{\sigma}}) \cdot \mathbf{U}_\theta^{[j]} + b_{A,j}^{(a)}\right], \tag{13}$$

$$\frac{\partial \log[\rho_{\boldsymbol{v}}(\boldsymbol{\sigma}, \tilde{\boldsymbol{\sigma}})]}{\partial \mathbf{U}_\theta^{[j]}} = \frac{i}{2}\,(\boldsymbol{\sigma} - \tilde{\boldsymbol{\sigma}})\,\frac{\mathcal{G}'}{\mathcal{G}}\left[\frac{1}{2}(\boldsymbol{\sigma} + \tilde{\boldsymbol{\sigma}}) \cdot \mathbf{U}_A^{[j]} + \frac{i}{2}(\boldsymbol{\sigma} - \tilde{\boldsymbol{\sigma}}) \cdot \mathbf{U}_\theta^{[j]} + b_{A,j}^{(a)}\right], \tag{14}$$

$$\tag{15}$$

where $\mathcal{G}'(x) = -e^{-x}$ and $\frac{\mathcal{G}'}{\mathcal{G}}[x] = (e^x - 1)^{-1}$.

### III. SAMPLING OF THE COST FUNCTION

In the main text we show that the cost function $\mathcal{C}(\boldsymbol{v})$ defined in Eq. (7)

$$\mathcal{C}(\boldsymbol{v}) = \frac{\|d\hat{\rho}_{\boldsymbol{v}}/dt\|_2^2}{\|\hat{\rho}_{\boldsymbol{v}}\|_2^2} = \frac{\operatorname{Tr}\left[\hat{\rho}_{\boldsymbol{v}}^\dagger \mathcal{L}^\dagger \mathcal{L} \hat{\rho}_{\boldsymbol{v}}\right]}{\operatorname{Tr}\left[\hat{\rho}_{\boldsymbol{v}}^\dagger \hat{\rho}_{\boldsymbol{v}}\right]}, \tag{16}$$

can be rewritten in such a way that it is possible to sample it from a distribution $p_{\boldsymbol{v}}(\sigma, \tilde{\sigma})$. In particoular, it can be rewritten in two different forms:

$$\mathcal{C}(\boldsymbol{v}) = \sum_{\boldsymbol{\sigma}, \tilde{\boldsymbol{\sigma}}} p_{\boldsymbol{v}}(\boldsymbol{\sigma}, \tilde{\boldsymbol{\sigma}}) \left( \sum_{\boldsymbol{\sigma}', \tilde{\boldsymbol{\sigma}}'} (\mathcal{L}^\dagger \mathcal{L})(\boldsymbol{\sigma}, \tilde{\boldsymbol{\sigma}}; \boldsymbol{\sigma}', \tilde{\boldsymbol{\sigma}}') \frac{\rho_{\boldsymbol{v}}(\boldsymbol{\sigma}', \tilde{\boldsymbol{\sigma}}')}{\rho_{\boldsymbol{v}}(\boldsymbol{\sigma}, \tilde{\boldsymbol{\sigma}})} \right), \tag{17}$$

or

$$\mathcal{C}(\boldsymbol{v}) = \sum_{\boldsymbol{\sigma}, \tilde{\boldsymbol{\sigma}}} p_{\boldsymbol{v}}(\boldsymbol{\sigma}, \tilde{\boldsymbol{\sigma}}) \left| \sum_{\boldsymbol{\sigma}', \tilde{\boldsymbol{\sigma}}'} \mathcal{L}(\boldsymbol{\sigma}, \tilde{\boldsymbol{\sigma}}; \boldsymbol{\sigma}', \tilde{\boldsymbol{\sigma}}') \frac{\rho_{\boldsymbol{v}}(\boldsymbol{\sigma}', \tilde{\boldsymbol{\sigma}}')}{\rho_{\boldsymbol{v}}(\boldsymbol{\sigma}, \tilde{\boldsymbol{\sigma}})} \right|^2. \tag{18}$$

In the main text we only considered the latter, but Eq. (17) and Eq. (18) are algebraically equivalent. This equivalence is only valid as long as the sum $\sum_{\boldsymbol{\sigma}, \tilde{\boldsymbol{\sigma}}}$ is performed exactly over the whole space $\mathcal{H} \otimes \mathcal{H}$. However, when the sum is evaluated via a Monte Carlo procedure, (17) and Eq. (18) have different convergence properties. In particular, Eq. (17) can be interpreted as the energy of the state $\rho_{\boldsymbol{v}}$ according to the *Hamiltonian* $\mathcal{L}^\dagger \mathcal{L}$, where every term in the sum can have either positive or negative real part. Equation (18), instead, is a sum over positive real terms which are respect the 0-variance property for the steady-state. This leads to a faster convergence of Eq. 18 and improved numerical stability when $\mathcal{C}(\boldsymbol{v})$ is evaluated through a Markov Chain. This is similar to what reported in several works that compare the performance of straightforward energy minimisation schemes to that of variance minimisation [2, 3].

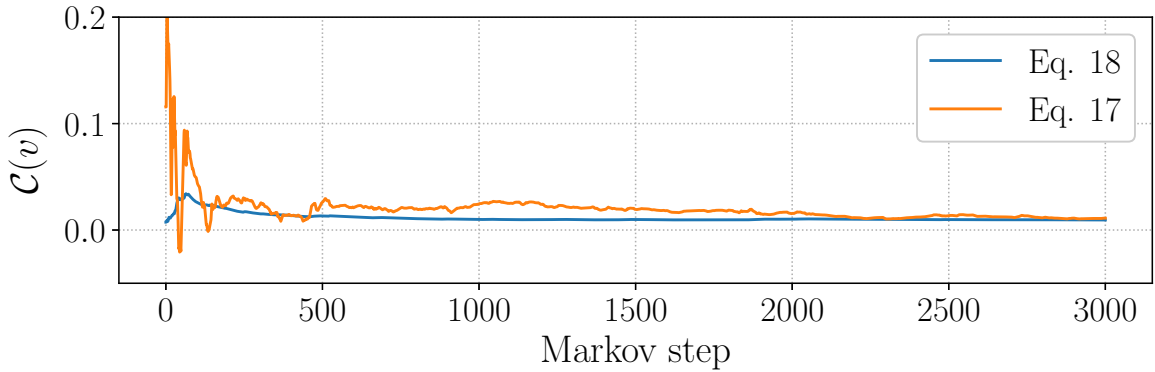

FIG. 1: The cost function Eq. 16 computed iteratively with the two schemes Eq. 17 and Eq. 18 along the same chain. It is clear that the latter converges much faster to the expected results.

To give an explicit example, in Fig. 1 the cost function for a set of weights far from the steady-state has been estimated both with Eq.(17) and Eq.(18) along a sample chain of length $M = 3000$. From the point of view of the computational cost, when computing the local terms $\mathcal{C}^{\text{Loc}}(\boldsymbol{\sigma}, \tilde{\boldsymbol{\sigma}})$ one must evaluate the neural network once for every pair $(\boldsymbol{\sigma}', \tilde{\boldsymbol{\sigma}}')$ for which the tensors $(\mathcal{L}^\dagger \mathcal{L})(\boldsymbol{\sigma}, \tilde{\boldsymbol{\sigma}}; \boldsymbol{\sigma}', \tilde{\boldsymbol{\sigma}}')$ or $\mathcal{L}(\boldsymbol{\sigma}, \tilde{\boldsymbol{\sigma}}; \boldsymbol{\sigma}', \tilde{\boldsymbol{\sigma}}')$ are non-zero. Clearly, $\mathcal{L}^\dagger \mathcal{L}$ is less sparse than $\mathcal{L}$, as it has approximately squared the number of non-zero entries, therefore it is computationally more expensive than estimating Eq. 18.

## IV. STOCHASTIC RECONFIGURATION

To improve convergence toward the global minimum of the cost function, we adopt the stochastic reconfiguration method, also known as natural gradient by the machine learning community [4, 5]. The stochastic reconfiguration corrects the gradient descent so that small variations $\boldsymbol{\delta v}$ in the parameters do not cause large variations, according to a certain norm, for the density matrices. The generalization of this method, previously developed for closed systems [6], can be straightforwardly generalized to our case. In what follows we give a sketch of the derivation.

Consider the first-order Taylor expansion of the variational density matrix with respect to the variational parameters:

$$\hat{\rho}_{\boldsymbol{v}+\boldsymbol{\delta v}} = \hat{\rho}_{\boldsymbol{v}} + \sum_i \delta v_i \mathcal{O}_{v_i} \hat{\rho}_{\boldsymbol{v}}, \tag{19}$$

where $v_i$, with $i \in [1, N_{\text{par}}]$ spans all the variational parameters. The distance between $\hat{\rho}_{\boldsymbol{v}}$ and $\hat{\rho}_{\boldsymbol{v}+\boldsymbol{\delta v}}$ is defined to be:

$$\delta s_{\boldsymbol{v}}^2 = \left\| \frac{\hat{\rho}_{\boldsymbol{v}+\boldsymbol{\delta v}}}{\|\hat{\rho}_{\boldsymbol{v}+\boldsymbol{\delta v}}\|} - \frac{\hat{\rho}_{\boldsymbol{v}}}{\|\hat{\rho}_{\boldsymbol{v}}\|} \right\|^2, \tag{20}$$

where in the following we have chosen the 2-norm, defined by $\|\hat{A}\|_2^2 = \text{Tr}\left[\hat{A}^\dagger \hat{A}\right]$. It is also known that the distance between two vectors in a space is given by the metric tensor $S$ as

$$\delta s_{\boldsymbol{v}}^2 = \sum_{i,j} \delta v_i S_{\boldsymbol{v}}^{i,j} \delta v_j. \tag{21}$$

After some algebra and keeping only the leading terms of the expansion, it is possible to write the (local) metric tensor as a quantity that can be easily sampled, namely:

$$S_{\boldsymbol{v}}^{i,j} = \sum_{\boldsymbol{\sigma}, \tilde{\boldsymbol{\sigma}}} p(\boldsymbol{\sigma}, \tilde{\boldsymbol{\sigma}}) \mathcal{O}_{\boldsymbol{v}}^i(\boldsymbol{\sigma}, \tilde{\boldsymbol{\sigma}}) \mathcal{O}_{\boldsymbol{v}}^j(\boldsymbol{\sigma}, \tilde{\boldsymbol{\sigma}}) - \left( \sum_{\boldsymbol{\sigma}, \tilde{\boldsymbol{\sigma}}} p(\boldsymbol{\sigma}, \tilde{\boldsymbol{\sigma}}) \mathcal{O}_{\boldsymbol{v}}^i(\boldsymbol{\sigma}, \tilde{\boldsymbol{\sigma}}) \right) \left( \sum_{\boldsymbol{\sigma}, \tilde{\boldsymbol{\sigma}}} p(\boldsymbol{\sigma}, \tilde{\boldsymbol{\sigma}}) \mathcal{O}_{\boldsymbol{v}}^j(\boldsymbol{\sigma}, \tilde{\boldsymbol{\sigma}}) \right). \tag{22}$$

Here, we also provide the explicit expressions of the components of the gradient of the cost function defined in Eq. (10) of the main text:

$$\nabla^i_{\boldsymbol{v}} \mathcal{C}(\boldsymbol{v}) = \sum_{\boldsymbol{\sigma},\tilde{\boldsymbol{\sigma}}} p_{\boldsymbol{v}}(\boldsymbol{\sigma},\tilde{\boldsymbol{\sigma}}) \mathcal{C}^{\mathrm{Loc}}(\boldsymbol{v},\boldsymbol{\sigma},\tilde{\boldsymbol{\sigma}})^{\star} \Big( \sum_{\boldsymbol{\sigma}',\tilde{\boldsymbol{\sigma}}'} \mathcal{L}(\boldsymbol{\sigma},\tilde{\boldsymbol{\sigma}};\boldsymbol{\sigma}',\tilde{\boldsymbol{\sigma}}') \frac{\rho_{\boldsymbol{v}}(\boldsymbol{\sigma}',\tilde{\boldsymbol{\sigma}}')}{\rho_{\boldsymbol{v}}(\boldsymbol{\sigma},\tilde{\boldsymbol{\sigma}})} \mathcal{O}_{\boldsymbol{v}}(\boldsymbol{\sigma}',\tilde{\boldsymbol{\sigma}}') \Big) -$$
$$- \left( \sum_{\boldsymbol{\sigma},\tilde{\boldsymbol{\sigma}}} p(\boldsymbol{\sigma},\tilde{\boldsymbol{\sigma}}) |\mathcal{C}^{\mathrm{Loc}}(\boldsymbol{v},\boldsymbol{\sigma},\tilde{\boldsymbol{\sigma}})|^2 \right) \left( \sum_{\boldsymbol{\sigma},\tilde{\boldsymbol{\sigma}}} p(\boldsymbol{\sigma},\tilde{\boldsymbol{\sigma}}) \mathcal{O}^i_{\boldsymbol{v}}(\boldsymbol{\sigma},\tilde{\boldsymbol{\sigma}}) \right). \quad (23)$$

In a standard gradient descent scheme one would update the variational parameters at each iteration according to the steepest gradient:

$$v_i \to v_i + \delta v_i \quad \text{where} \quad \delta v_i = -\nabla^i_{\boldsymbol{v}} \mathcal{C}(\boldsymbol{v}). \quad (24)$$

Instead, if we also require that $\delta s^2$ be small, we obtain that $\boldsymbol{\delta v}$ is determined by the linear system of equations

$$S^{i,j}_{\boldsymbol{v}} \delta v_j = \nabla^i_{\boldsymbol{v}} \mathcal{C}(\boldsymbol{v}) \quad (25)$$

where $S_{\boldsymbol{v}}$ is a real, dense, symmetric matrix of size $N_{\mathrm{par}} \times N_{\mathrm{par}}$. $S$ has often many degenerate eigenvalues. As in this work $N_{\mathrm{par}} > 1000$, Krylov-space iterative solvers such as MINRES-QLP are particularly suitable [7]. To improve the stability of the solution we solved the equivalent system

$$(S^{i,j}_{\boldsymbol{v}} + b\,\delta_{i,j}) \delta v_j = \nabla^i_{\boldsymbol{v}} \mathcal{C}(\boldsymbol{v}) \quad (26)$$

with $b \in [10^{-4}, 10^{-3}]$.

## V. NOTES ON THE NUMERICS

The core of the algorithm consists in sampling with a Markov-Chain Monte Carlo the two quantities Eq.(20) and Eq.(21), and then solving the linear system defined by Eq.(23). The Monte Carlo step of the algorithm can be easily and efficiently distributed across many cores. Unfortunately, to the best of our knowledge there do not exist efficient distributed solvers for dense systems of linear equations, so the second part of the algorithm cannot be parallelized as efficiently (although the speedup from performing matrix-vector multiplications on several threads is still significant). Therefore the bottleneck of the algorithm is, as of now, the numerical solution of Eq. 23.

We wrote our code in Julia, using MPI.jl to distribute the Monte Carlo Chains across several processors and/or nodes. The linear system of equations was solved either by inversion with standard library routines or by linking to the FORTRAN routines of MINRES-QLP developed by the autors of [7].

## VI. CPU TIME

In Fig. 2 we report the scaling of the runtime performance of the neural network algorithm, compared to traditional brute-force Runge-Kutta integration of the master equation and to a quantum trajectory approach.

## VII. PROOF OF PRINCIPLE FOR 2D SYSTEMS

In this section we show how the neural ansatz presented in the main text is also suitable to describe the steady-state of 2D lattices. We consider again the dissipative quantum Ising model [see Eq. (1) and Eq. (12) in the main text] for a $3 \times 3$ lattice with periodic boundary conditions. For $V/\gamma = 2$ and different values of the transverse magnetic field we find again a good agreement with the exact results (obtained via brute-force integration of the master equation). A summary of the results is shown in Table I.

---

[1] J. Cui, J. I. Cirac, and M. C. Bañuls, Phys. Rev. Lett. **114**, 220601 (2015).

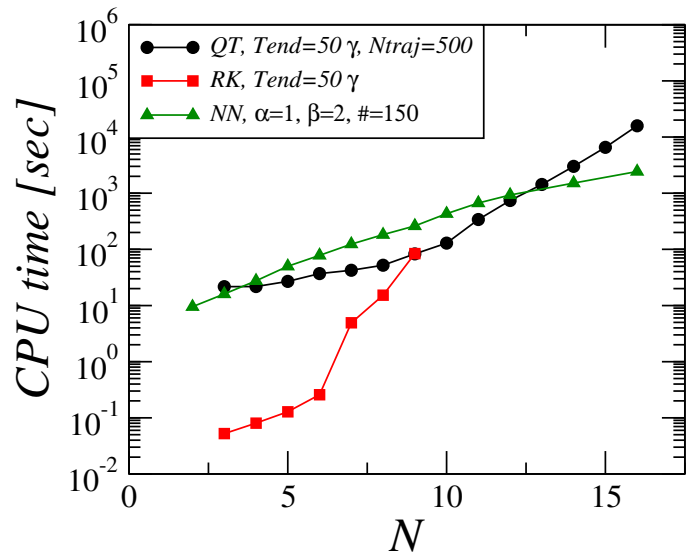

FIG. 2: Runtime cost (in CPU-core-seconds) of various algorithms as a function of the system size $N$. Red squared: the brute-force RK45 integration of the master equation up to $T_{\text{end}} = 50\gamma$. Green triangles: the integration of 500 trajectories up to $T_{\text{end}} = 50\gamma$ with the TSIT5 ode-integrator. Black circles: the CPU-time of the Neural Network algorithm presented in this paper, with $\alpha = 1$, $\beta = 2$, $M = 10^3 * (1 + log(N))$ Markov-chain samples and 150 iterations.

| $g/\gamma$ | $\langle\sigma_x\rangle$ | $\epsilon_{\text{rel}}[\langle\sigma_x\rangle]$ | $\langle\sigma_y\rangle$ | $\epsilon_{\text{rel}}[\langle\sigma_y\rangle]$ | $\langle\sigma_z\rangle$ | $\epsilon_{\text{rel}}[\langle\sigma_z\rangle]$ |
|---|---|---|---|---|---|---|
| 0.5 | 0.123 | 0.007 % | 0.0158 | 0.979 % | -0.992 | 0.051 % |
| 0.8 | 0.198 | 0.104 % | 0.0258 | 0.598 % | -0.978 | 0.103 % |
| 1.1 | 0.274 | 0.134 % | 0.0370 | 0.529 % | -0.959 | 0.061 % |
| 1.4 | 0.350 | 0.117 % | 0.0506 | 0.495 % | -0.929 | 0.007 % |
| 1.7 | 0.427 | 0.101 % | 0.0672 | 2.154 % | -0.888 | 0.522 % |

TABLE I: Steady-state values of local magnetizations $\langle\sigma_\alpha\rangle$ with $\{\alpha = x, y, z\}$ for a $3 \times 3$ lattice with periodic boundary conditions for $V/\gamma = 2$. The relative error with respect to the exact solution is also shown.

[2] C. J. Umrigar and C. Filippi, Physical Review Letters **94** (2005), 10.1103/physrevlett.94.150201.
[3] P. R. C. Kent, R. J. Needs, and G. Rajagopal, Physical Review B **59**, 12344 (1999).
[4] R. Pascanu and Y. Bengio, (2013), arXiv:arXiv:1301.3584 [cs.LG].
[5] J. Martens, (2014), arXiv:1412.1193 [cs.LG].
[6] F. Becca and S. Sorella, in *Quantum Monte Carlo Approaches Correl. Syst.* (Cambridge University Press, 2017) pp. 131–155.
[7] S.-C. T. Choi and M. A. Saunders, ACM Transactions on Mathematical Software **40**, 1 (2014).



\end{document}

%% file: NDM.tex
\begin{tikzpicture}[scale=2]
    \draw[fill=colmid, rounded corners] (-0.2, -0.2) rectangle (3.7, 0.2) {};
    \draw[fill=colup, rounded corners] (0.3, 0.8) rectangle (3.2, 1.2) {};
    \draw[fill=coldown, rounded corners] (0.3, -1.2) rectangle (3.2, -0.8) {};

    \node (v1)[neuron] at (0, 0) {$\sigma_1$};
    \node (v2)[neuron] at (1, 0) {$\sigma_2$};
    \node (v3)[neuron] at (2, 0) {$\sigma_3$};
    \node (vm) at (2.75, 0) {$\dots$};
    \node (v4)[neuron] at (3.5, 0) {$\sigma_N$};
    \node[right=0.1cm of v4] (v) {$\boldsymbol{\sigma} \in \{0, 1\}^{N}$};
    \node[learned,below=0.03cm of v1] (bv1) {$b^{(\sigma)}_{\nu,1}$};
    \node[learned,below=0.03cm of v4] (bv4) {$b^{(\sigma)}_{\nu,N}$};

    \node (h1)[neuron] at (0.5, 1) {$h_1$};
    \node (h2)[neuron] at (1.5, 1) {$h_2$};
    \node (hh)		   at (2.25, 1) {$\dots$};
    \node (h3)[neuron] at (3.0, 1) {$h_{N_{h}}$};
    \node[right=0.1cm of h3] (h) {$\textbf{h} \in \{0, 1\}^{N_h}$};
    \node[learned,above=0.1cm of h1] (bh1) {$b^{(h)}_{\nu,1}$};
    \node[learned,above=0.1cm of h2] (bh2) {$b^{(h)}_{\nu,2}$};
    \node[learned,above=0.1cm of h3] (bh3) {$b^{(h)}_{\nu,N_h}$};

	\node (a1)[neuron] at (0.5, -1) {$a_1$};
	\node (a2)[neuron] at (1.5, -1) {$a_2$};
	\node (aa)		   at (2.25, -1) {$\dots$};
	\node (a3)[neuron] at (3.0, -1) {$a_{N_{a}}$};
    \node[right=0.1cm of a3] (a) {$\textbf{a} \in \{0, 1\}^{N_a}$};
    \node[learned,below=0.1cm of a1] (ba1) {$b^{(a)}_{\nu,1}$};
    \node[learned,below=0.1cm of a2] (ba2) {$b^{(a)}_{\nu,2}$};
    \node[learned,below=0.1cm of a3] (ba3) {$b^{(a)}_{\nu,N_{a}}$};

    \node[learned] (W) at (4.05, 0.55) {$W_\nu \in \mathbb{R}^{N_h \times N}$};
    \node[learned] (U) at (4.05, -0.55) {$U_\nu \in \mathbb{R}^{N_a \times N}$};

    \draw[learned,stateTransition] (v1) -- (h1) node [midway,above=-0.06cm,sloped] {$w_{\nu,(1,1)}$};
    \draw[stateTransition] (v1) -- (h2) node [midway,above=-0.06cm,sloped] {};
    \draw[stateTransition] (v1) -- (h3) node [midway,above=-0.06cm,sloped] {};

    \draw[stateTransition] (v2) -- (h1) node [midway,above=-0.06cm,sloped] {};
    \draw[stateTransition] (v2) -- (h2) node [midway,above=-0.06cm,sloped] {};
    \draw[stateTransition] (v2) -- (h3) node [midway,above=-0.06cm,sloped] {};

    \draw[stateTransition] (v3) -- (h1) node [midway,above=-0.06cm,sloped] {};
    \draw[stateTransition] (v3) -- (h2) node [midway,above=-0.06cm,sloped] {};
    \draw[stateTransition] (v3) -- (h3) node [midway,above=-0.06cm,sloped] {};

    \draw[stateTransition] (v4) -- (h1) node [midway,above=-0.06cm,sloped] {};
    \draw[stateTransition] (v4) -- (h2) node [midway,above=-0.06cm,sloped] {};
    \draw[learned,stateTransition] (v4) -- (h3) node [midway,above=-0.06cm,sloped] {$w_{\nu,(4,3)}$};
    
    \draw[learned,stateTransition] (v1) -- (a1) node [midway, above=-0.06cm, sloped] {$u_{\nu,(1,1)}$};
    \draw[stateTransition] (v1) -- (a2) node [midway, above=-0.06cm, sloped] {};
    \draw[stateTransition] (v1) -- (a3) node [midway, above=-0.06cm, sloped] {};
    \draw[stateTransition] (v2) -- (a1) node [midway, above=-0.06cm, sloped] {};
    \draw[stateTransition] (v2) -- (a2) node [midway, above=-0.06cm, sloped] {};
    \draw[stateTransition] (v2) -- (a3) node [midway, above=-0.06cm, sloped] {};
    \draw[stateTransition] (v3) -- (a1) node [midway, above=-0.06cm, sloped] {};
    \draw[stateTransition] (v3) -- (a2) node [midway, above=-0.06cm, sloped] {};
    \draw[stateTransition] (v3) -- (a3) node [midway, above=-0.06cm, sloped] {};
    \draw[stateTransition] (v4) -- (a1) node [midway, above=-0.06cm, sloped] {};
    \draw[stateTransition] (v4) -- (a2) node [midway, above=-0.06cm, sloped] {};
    \draw[learned,stateTransition] (v4) -- (a3) node [midway, above=-0.06cm, sloped] {$u_{\nu,(4,3)}$};

\end{tikzpicture}

%% file: arXiv_v2.bbl
\begin{thebibliography}{57}%
\makeatletter
\providecommand \@ifxundefined [1]{%
 \@ifx{#1\undefined}
}%
\providecommand \@ifnum [1]{%
 \ifnum #1\expandafter \@firstoftwo
 \else \expandafter \@secondoftwo
 \fi
}%
\providecommand \@ifx [1]{%
 \ifx #1\expandafter \@firstoftwo
 \else \expandafter \@secondoftwo
 \fi
}%
\providecommand \natexlab [1]{#1}%
\providecommand \enquote  [1]{``#1''}%
\providecommand \bibnamefont  [1]{#1}%
\providecommand \bibfnamefont [1]{#1}%
\providecommand \citenamefont [1]{#1}%
\providecommand \href@noop [0]{\@secondoftwo}%
\providecommand \href [0]{\begingroup \@sanitize@url \@href}%
\providecommand \@href[1]{\@@startlink{#1}\@@href}%
\providecommand \@@href[1]{\endgroup#1\@@endlink}%
\providecommand \@sanitize@url [0]{\catcode `\\12\catcode `\$12\catcode
  `\&12\catcode `\#12\catcode `\^12\catcode `\_12\catcode `\%12\relax}%
\providecommand \@@startlink[1]{}%
\providecommand \@@endlink[0]{}%
\providecommand \url  [0]{\begingroup\@sanitize@url \@url }%
\providecommand \@url [1]{\endgroup\@href {#1}{\urlprefix }}%
\providecommand \urlprefix  [0]{URL }%
\providecommand \Eprint [0]{\href }%
\providecommand \doibase [0]{http://dx.doi.org/}%
\providecommand \selectlanguage [0]{\@gobble}%
\providecommand \bibinfo  [0]{\@secondoftwo}%
\providecommand \bibfield  [0]{\@secondoftwo}%
\providecommand \translation [1]{[#1]}%
\providecommand \BibitemOpen [0]{}%
\providecommand \bibitemStop [0]{}%
\providecommand \bibitemNoStop [0]{.\EOS\space}%
\providecommand \EOS [0]{\spacefactor3000\relax}%
\providecommand \BibitemShut  [1]{\csname bibitem#1\endcsname}%
\let\auto@bib@innerbib\@empty
\bibitem [{\citenamefont {Breuer}\ and\ \citenamefont
  {Petruccione}(2007)}]{Breuer2007}%
  \BibitemOpen
  \bibfield  {author} {\bibinfo {author} {\bibfnamefont {H.-P.}\ \bibnamefont
  {Breuer}}\ and\ \bibinfo {author} {\bibfnamefont {F.}~\bibnamefont
  {Petruccione}},\ }\href {\doibase 10.1093/acprof:oso/9780199213900.001.0001}
  {\emph {\bibinfo {title} {The Theory of Open Quantum Systems}}}\ (\bibinfo
  {publisher} {Oxford University Press},\ \bibinfo {year} {2007})\BibitemShut
  {NoStop}%
\bibitem [{\citenamefont {Devoret}\ and\ \citenamefont
  {Schoelkopf}(2013)}]{Devoret2013}%
  \BibitemOpen
  \bibfield  {author} {\bibinfo {author} {\bibfnamefont {M.~H.}\ \bibnamefont
  {Devoret}}\ and\ \bibinfo {author} {\bibfnamefont {R.~J.}\ \bibnamefont
  {Schoelkopf}},\ }\href {\doibase 10.1126/science.1231930} {\bibfield
  {journal} {\bibinfo  {journal} {Science}\ }\textbf {\bibinfo {volume}
  {339}},\ \bibinfo {pages} {1169} (\bibinfo {year} {2013})}\BibitemShut
  {NoStop}%
\bibitem [{\citenamefont {Carusotto}\ and\ \citenamefont
  {Ciuti}(2013)}]{CarusottoRMP13}%
  \BibitemOpen
  \bibfield  {author} {\bibinfo {author} {\bibfnamefont {I.}~\bibnamefont
  {Carusotto}}\ and\ \bibinfo {author} {\bibfnamefont {C.}~\bibnamefont
  {Ciuti}},\ }\href {\doibase 10.1103/RevModPhys.85.299} {\bibfield  {journal}
  {\bibinfo  {journal} {Rev. Mod. Phys.}\ }\textbf {\bibinfo {volume} {85}},\
  \bibinfo {pages} {299} (\bibinfo {year} {2013})}\BibitemShut {NoStop}%
\bibitem [{\citenamefont {Noh}\ and\ \citenamefont
  {Angelakis}(2016)}]{Noh2016}%
  \BibitemOpen
  \bibfield  {author} {\bibinfo {author} {\bibfnamefont {C.}~\bibnamefont
  {Noh}}\ and\ \bibinfo {author} {\bibfnamefont {D.~G.}\ \bibnamefont
  {Angelakis}},\ }\href {\doibase 10.1088/0034-4885/80/1/016401} {\bibfield
  {journal} {\bibinfo  {journal} {Reports on Progress in Physics}\ }\textbf
  {\bibinfo {volume} {80}},\ \bibinfo {pages} {016401} (\bibinfo {year}
  {2016})}\BibitemShut {NoStop}%
\bibitem [{\citenamefont {Hartmann}(2016)}]{Hartmann2016}%
  \BibitemOpen
  \bibfield  {author} {\bibinfo {author} {\bibfnamefont {M.~J.}\ \bibnamefont
  {Hartmann}},\ }\href {\doibase 10.1088/2040-8978/18/10/104005} {\bibfield
  {journal} {\bibinfo  {journal} {Journal of Optics}\ }\textbf {\bibinfo
  {volume} {18}},\ \bibinfo {pages} {104005} (\bibinfo {year}
  {2016})}\BibitemShut {NoStop}%
\bibitem [{\citenamefont {Verstraete}\ \emph {et~al.}(2009)\citenamefont
  {Verstraete}, \citenamefont {Wolf},\ and\ \citenamefont
  {Ignacio~Cirac}}]{Verstraete:2009fk}%
  \BibitemOpen
  \bibfield  {author} {\bibinfo {author} {\bibfnamefont {F.}~\bibnamefont
  {Verstraete}}, \bibinfo {author} {\bibfnamefont {M.~M.}\ \bibnamefont
  {Wolf}}, \ and\ \bibinfo {author} {\bibfnamefont {J.}~\bibnamefont
  {Ignacio~Cirac}},\ }\href {https://doi.org/10.1038/nphys1342} {\bibfield
  {journal} {\bibinfo  {journal} {Nature Physics}\ }\textbf {\bibinfo {volume}
  {5}},\ \bibinfo {pages} {633 EP } (\bibinfo {year} {2009})}\BibitemShut
  {NoStop}%
\bibitem [{\citenamefont {Barreiro}\ \emph {et~al.}(2011)\citenamefont
  {Barreiro}, \citenamefont {M{\"u}ller}, \citenamefont {Schindler},
  \citenamefont {Nigg}, \citenamefont {Monz}, \citenamefont {Chwalla},
  \citenamefont {Hennrich}, \citenamefont {Roos}, \citenamefont {Zoller},\ and\
  \citenamefont {Blatt}}]{Barreiro:2011fj}%
  \BibitemOpen
  \bibfield  {author} {\bibinfo {author} {\bibfnamefont {J.~T.}\ \bibnamefont
  {Barreiro}}, \bibinfo {author} {\bibfnamefont {M.}~\bibnamefont
  {M{\"u}ller}}, \bibinfo {author} {\bibfnamefont {P.}~\bibnamefont
  {Schindler}}, \bibinfo {author} {\bibfnamefont {D.}~\bibnamefont {Nigg}},
  \bibinfo {author} {\bibfnamefont {T.}~\bibnamefont {Monz}}, \bibinfo {author}
  {\bibfnamefont {M.}~\bibnamefont {Chwalla}}, \bibinfo {author} {\bibfnamefont
  {M.}~\bibnamefont {Hennrich}}, \bibinfo {author} {\bibfnamefont {C.~F.}\
  \bibnamefont {Roos}}, \bibinfo {author} {\bibfnamefont {P.}~\bibnamefont
  {Zoller}}, \ and\ \bibinfo {author} {\bibfnamefont {R.}~\bibnamefont
  {Blatt}},\ }\href {https://doi.org/10.1038/nature09801} {\bibfield  {journal}
  {\bibinfo  {journal} {Nature}\ }\textbf {\bibinfo {volume} {470}},\ \bibinfo
  {pages} {486 EP } (\bibinfo {year} {2011})}\BibitemShut {NoStop}%
\bibitem [{\citenamefont {Sieberer}\ \emph {et~al.}(2016)\citenamefont
  {Sieberer}, \citenamefont {Buchhold},\ and\ \citenamefont
  {Diehl}}]{Sieberer2016}%
  \BibitemOpen
  \bibfield  {author} {\bibinfo {author} {\bibfnamefont {L.~M.}\ \bibnamefont
  {Sieberer}}, \bibinfo {author} {\bibfnamefont {M.}~\bibnamefont {Buchhold}},
  \ and\ \bibinfo {author} {\bibfnamefont {S.}~\bibnamefont {Diehl}},\ }\href
  {http://stacks.iop.org/0034-4885/79/i=9/a=096001} {\bibfield  {journal}
  {\bibinfo  {journal} {Reports on Progress in Physics}\ }\textbf {\bibinfo
  {volume} {79}},\ \bibinfo {pages} {096001} (\bibinfo {year}
  {2016})}\BibitemShut {NoStop}%
\bibitem [{\citenamefont {Maghrebi}\ and\ \citenamefont
  {Gorshkov}(2016)}]{Maghrebi16}%
  \BibitemOpen
  \bibfield  {author} {\bibinfo {author} {\bibfnamefont {M.~F.}\ \bibnamefont
  {Maghrebi}}\ and\ \bibinfo {author} {\bibfnamefont {A.~V.}\ \bibnamefont
  {Gorshkov}},\ }\href {\doibase 10.1103/PhysRevB.93.014307} {\bibfield
  {journal} {\bibinfo  {journal} {Phys. Rev. B}\ }\textbf {\bibinfo {volume}
  {93}},\ \bibinfo {pages} {014307} (\bibinfo {year} {2016})}\BibitemShut
  {NoStop}%
\bibitem [{\citenamefont {Mascarenhas}\ \emph {et~al.}(2015)\citenamefont
  {Mascarenhas}, \citenamefont {Flayac},\ and\ \citenamefont
  {Savona}}]{MascarenhasPRA2015}%
  \BibitemOpen
  \bibfield  {author} {\bibinfo {author} {\bibfnamefont {E.}~\bibnamefont
  {Mascarenhas}}, \bibinfo {author} {\bibfnamefont {H.}~\bibnamefont {Flayac}},
  \ and\ \bibinfo {author} {\bibfnamefont {V.}~\bibnamefont {Savona}},\ }\href
  {\doibase 10.1103/PhysRevA.92.022116} {\bibfield  {journal} {\bibinfo
  {journal} {Phys. Rev. A}\ }\textbf {\bibinfo {volume} {92}},\ \bibinfo
  {pages} {022116} (\bibinfo {year} {2015})}\BibitemShut {NoStop}%
\bibitem [{\citenamefont {Cui}\ \emph {et~al.}(2015)\citenamefont {Cui},
  \citenamefont {Cirac},\ and\ \citenamefont {Ba\~nuls}}]{CuiPRL2015}%
  \BibitemOpen
  \bibfield  {author} {\bibinfo {author} {\bibfnamefont {J.}~\bibnamefont
  {Cui}}, \bibinfo {author} {\bibfnamefont {J.~I.}\ \bibnamefont {Cirac}}, \
  and\ \bibinfo {author} {\bibfnamefont {M.~C.}\ \bibnamefont {Ba\~nuls}},\
  }\href {\doibase 10.1103/PhysRevLett.114.220601} {\bibfield  {journal}
  {\bibinfo  {journal} {Phys. Rev. Lett.}\ }\textbf {\bibinfo {volume} {114}},\
  \bibinfo {pages} {220601} (\bibinfo {year} {2015})}\BibitemShut {NoStop}%
\bibitem [{\citenamefont {Jaschke}\ \emph {et~al.}(2019)\citenamefont
  {Jaschke}, \citenamefont {Montangero},\ and\ \citenamefont
  {Carr}}]{Jaschke2019}%
  \BibitemOpen
  \bibfield  {author} {\bibinfo {author} {\bibfnamefont {D.}~\bibnamefont
  {Jaschke}}, \bibinfo {author} {\bibfnamefont {S.}~\bibnamefont {Montangero}},
  \ and\ \bibinfo {author} {\bibfnamefont {L.~D.}\ \bibnamefont {Carr}},\
  }\href {http://stacks.iop.org/2058-9565/4/i=1/a=013001} {\bibfield  {journal}
  {\bibinfo  {journal} {Quantum Science and Technology}\ }\textbf {\bibinfo
  {volume} {4}},\ \bibinfo {pages} {013001} (\bibinfo {year}
  {2019})}\BibitemShut {NoStop}%
\bibitem [{\citenamefont {Werner}\ \emph {et~al.}(2016)\citenamefont {Werner},
  \citenamefont {Jaschke}, \citenamefont {Silvi}, \citenamefont {Kliesch},
  \citenamefont {Calarco}, \citenamefont {Eisert},\ and\ \citenamefont
  {Montangero}}]{WernerPRL2016}%
  \BibitemOpen
  \bibfield  {author} {\bibinfo {author} {\bibfnamefont {A.~H.}\ \bibnamefont
  {Werner}}, \bibinfo {author} {\bibfnamefont {D.}~\bibnamefont {Jaschke}},
  \bibinfo {author} {\bibfnamefont {P.}~\bibnamefont {Silvi}}, \bibinfo
  {author} {\bibfnamefont {M.}~\bibnamefont {Kliesch}}, \bibinfo {author}
  {\bibfnamefont {T.}~\bibnamefont {Calarco}}, \bibinfo {author} {\bibfnamefont
  {J.}~\bibnamefont {Eisert}}, \ and\ \bibinfo {author} {\bibfnamefont
  {S.}~\bibnamefont {Montangero}},\ }\href {\doibase
  10.1103/PhysRevLett.116.237201} {\bibfield  {journal} {\bibinfo  {journal}
  {Phys. Rev. Lett.}\ }\textbf {\bibinfo {volume} {116}},\ \bibinfo {pages}
  {237201} (\bibinfo {year} {2016})}\BibitemShut {NoStop}%
\bibitem [{\citenamefont {Kshetrimayum}\ \emph {et~al.}(2017)\citenamefont
  {Kshetrimayum}, \citenamefont {Weimer},\ and\ \citenamefont
  {Or\'{u}s}}]{OrusNat17}%
  \BibitemOpen
  \bibfield  {author} {\bibinfo {author} {\bibfnamefont {A.}~\bibnamefont
  {Kshetrimayum}}, \bibinfo {author} {\bibfnamefont {H.}~\bibnamefont
  {Weimer}}, \ and\ \bibinfo {author} {\bibfnamefont {R.}~\bibnamefont
  {Or\'{u}s}},\ }\href {\doibase 10.1038/s41467-017-01511-6} {\bibfield
  {journal} {\bibinfo  {journal} {Nature Communications}\ }\textbf {\bibinfo
  {volume} {8}},\ \bibinfo {pages} {1291} (\bibinfo {year} {2017})}\BibitemShut
  {NoStop}%
\bibitem [{\citenamefont {Biella}\ \emph {et~al.}(2018)\citenamefont {Biella},
  \citenamefont {Jin}, \citenamefont {Viyuela}, \citenamefont {Ciuti},
  \citenamefont {Fazio},\ and\ \citenamefont {Rossini}}]{BiellaNLCE2018}%
  \BibitemOpen
  \bibfield  {author} {\bibinfo {author} {\bibfnamefont {A.}~\bibnamefont
  {Biella}}, \bibinfo {author} {\bibfnamefont {J.}~\bibnamefont {Jin}},
  \bibinfo {author} {\bibfnamefont {O.}~\bibnamefont {Viyuela}}, \bibinfo
  {author} {\bibfnamefont {C.}~\bibnamefont {Ciuti}}, \bibinfo {author}
  {\bibfnamefont {R.}~\bibnamefont {Fazio}}, \ and\ \bibinfo {author}
  {\bibfnamefont {D.}~\bibnamefont {Rossini}},\ }\href {\doibase
  10.1103/PhysRevB.97.035103} {\bibfield  {journal} {\bibinfo  {journal} {Phys.
  Rev. B}\ }\textbf {\bibinfo {volume} {97}},\ \bibinfo {pages} {035103}
  (\bibinfo {year} {2018})}\BibitemShut {NoStop}%
\bibitem [{\citenamefont {Jin}\ \emph {et~al.}(2016)\citenamefont {Jin},
  \citenamefont {Biella}, \citenamefont {Viyuela}, \citenamefont {Mazza},
  \citenamefont {Keeling}, \citenamefont {Fazio},\ and\ \citenamefont
  {Rossini}}]{Jin16}%
  \BibitemOpen
  \bibfield  {author} {\bibinfo {author} {\bibfnamefont {J.}~\bibnamefont
  {Jin}}, \bibinfo {author} {\bibfnamefont {A.}~\bibnamefont {Biella}},
  \bibinfo {author} {\bibfnamefont {O.}~\bibnamefont {Viyuela}}, \bibinfo
  {author} {\bibfnamefont {L.}~\bibnamefont {Mazza}}, \bibinfo {author}
  {\bibfnamefont {J.}~\bibnamefont {Keeling}}, \bibinfo {author} {\bibfnamefont
  {R.}~\bibnamefont {Fazio}}, \ and\ \bibinfo {author} {\bibfnamefont
  {D.}~\bibnamefont {Rossini}},\ }\href {\doibase 10.1103/PhysRevX.6.031011}
  {\bibfield  {journal} {\bibinfo  {journal} {Phys. Rev. X}\ }\textbf {\bibinfo
  {volume} {6}},\ \bibinfo {pages} {031011} (\bibinfo {year}
  {2016})}\BibitemShut {NoStop}%
\bibitem [{\citenamefont {Finazzi}\ \emph {et~al.}(2015)\citenamefont
  {Finazzi}, \citenamefont {Le~Boit\'e}, \citenamefont {Storme}, \citenamefont
  {Baksic},\ and\ \citenamefont {Ciuti}}]{FinazziPRL15}%
  \BibitemOpen
  \bibfield  {author} {\bibinfo {author} {\bibfnamefont {S.}~\bibnamefont
  {Finazzi}}, \bibinfo {author} {\bibfnamefont {A.}~\bibnamefont {Le~Boit\'e}},
  \bibinfo {author} {\bibfnamefont {F.}~\bibnamefont {Storme}}, \bibinfo
  {author} {\bibfnamefont {A.}~\bibnamefont {Baksic}}, \ and\ \bibinfo {author}
  {\bibfnamefont {C.}~\bibnamefont {Ciuti}},\ }\href {\doibase
  10.1103/PhysRevLett.115.080604} {\bibfield  {journal} {\bibinfo  {journal}
  {Phys. Rev. Lett.}\ }\textbf {\bibinfo {volume} {115}},\ \bibinfo {pages}
  {080604} (\bibinfo {year} {2015})}\BibitemShut {NoStop}%
\bibitem [{\citenamefont {Rota}\ \emph {et~al.}(2017)\citenamefont {Rota},
  \citenamefont {Storme}, \citenamefont {Bartolo}, \citenamefont {Fazio},\ and\
  \citenamefont {Ciuti}}]{Rota17}%
  \BibitemOpen
  \bibfield  {author} {\bibinfo {author} {\bibfnamefont {R.}~\bibnamefont
  {Rota}}, \bibinfo {author} {\bibfnamefont {F.}~\bibnamefont {Storme}},
  \bibinfo {author} {\bibfnamefont {N.}~\bibnamefont {Bartolo}}, \bibinfo
  {author} {\bibfnamefont {R.}~\bibnamefont {Fazio}}, \ and\ \bibinfo {author}
  {\bibfnamefont {C.}~\bibnamefont {Ciuti}},\ }\href {\doibase
  10.1103/PhysRevB.95.134431} {\bibfield  {journal} {\bibinfo  {journal} {Phys.
  Rev. B}\ }\textbf {\bibinfo {volume} {95}},\ \bibinfo {pages} {134431}
  (\bibinfo {year} {2017})}\BibitemShut {NoStop}%
\bibitem [{\citenamefont {Rota}\ \emph {et~al.}(2018)\citenamefont {Rota},
  \citenamefont {Minganti}, \citenamefont {Ciuti},\ and\ \citenamefont
  {Savona}}]{RotaArx2018}%
  \BibitemOpen
  \bibfield  {author} {\bibinfo {author} {\bibfnamefont {R.}~\bibnamefont
  {Rota}}, \bibinfo {author} {\bibfnamefont {F.}~\bibnamefont {Minganti}},
  \bibinfo {author} {\bibfnamefont {C.}~\bibnamefont {Ciuti}}, \ and\ \bibinfo
  {author} {\bibfnamefont {V.}~\bibnamefont {Savona}},\ }\href@noop {} {\
  (\bibinfo {year} {2018})},\ \Eprint {http://arxiv.org/abs/1809.10138}
  {arXiv:1809.10138 [quant-ph]} \BibitemShut {NoStop}%
\bibitem [{\citenamefont {Casteels}\ \emph {et~al.}(2018)\citenamefont
  {Casteels}, \citenamefont {Wilson},\ and\ \citenamefont {Wouters}}]{Wim2018}%
  \BibitemOpen
  \bibfield  {author} {\bibinfo {author} {\bibfnamefont {W.}~\bibnamefont
  {Casteels}}, \bibinfo {author} {\bibfnamefont {R.~M.}\ \bibnamefont
  {Wilson}}, \ and\ \bibinfo {author} {\bibfnamefont {M.}~\bibnamefont
  {Wouters}},\ }\href {\doibase 10.1103/PhysRevA.97.062107} {\bibfield
  {journal} {\bibinfo  {journal} {Phys. Rev. A}\ }\textbf {\bibinfo {volume}
  {97}},\ \bibinfo {pages} {062107} (\bibinfo {year} {2018})}\BibitemShut
  {NoStop}%
\bibitem [{\citenamefont {Nagy}\ and\ \citenamefont
  {Savona}(2018)}]{NagyPRA18}%
  \BibitemOpen
  \bibfield  {author} {\bibinfo {author} {\bibfnamefont {A.}~\bibnamefont
  {Nagy}}\ and\ \bibinfo {author} {\bibfnamefont {V.}~\bibnamefont {Savona}},\
  }\href {\doibase 10.1103/PhysRevA.97.052129} {\bibfield  {journal} {\bibinfo
  {journal} {Phys. Rev. A}\ }\textbf {\bibinfo {volume} {97}},\ \bibinfo
  {pages} {052129} (\bibinfo {year} {2018})}\BibitemShut {NoStop}%
\bibitem [{\citenamefont {Shammah}\ \emph {et~al.}(2018)\citenamefont
  {Shammah}, \citenamefont {Ahmed}, \citenamefont {Lambert}, \citenamefont
  {De~Liberato},\ and\ \citenamefont {Nori}}]{ShammahPRA2018}%
  \BibitemOpen
  \bibfield  {author} {\bibinfo {author} {\bibfnamefont {N.}~\bibnamefont
  {Shammah}}, \bibinfo {author} {\bibfnamefont {S.}~\bibnamefont {Ahmed}},
  \bibinfo {author} {\bibfnamefont {N.}~\bibnamefont {Lambert}}, \bibinfo
  {author} {\bibfnamefont {S.}~\bibnamefont {De~Liberato}}, \ and\ \bibinfo
  {author} {\bibfnamefont {F.}~\bibnamefont {Nori}},\ }\href {\doibase
  10.1103/PhysRevA.98.063815} {\bibfield  {journal} {\bibinfo  {journal} {Phys.
  Rev. A}\ }\textbf {\bibinfo {volume} {98}},\ \bibinfo {pages} {063815}
  (\bibinfo {year} {2018})}\BibitemShut {NoStop}%
\bibitem [{\citenamefont {Vicentini}\ \emph {et~al.}(2018)\citenamefont
  {Vicentini}, \citenamefont {Minganti}, \citenamefont {Biella}, \citenamefont
  {Orso},\ and\ \citenamefont {Ciuti}}]{Vicentini18ArXiV}%
  \BibitemOpen
  \bibfield  {author} {\bibinfo {author} {\bibfnamefont {F.}~\bibnamefont
  {Vicentini}}, \bibinfo {author} {\bibfnamefont {F.}~\bibnamefont {Minganti}},
  \bibinfo {author} {\bibfnamefont {A.}~\bibnamefont {Biella}}, \bibinfo
  {author} {\bibfnamefont {G.}~\bibnamefont {Orso}}, \ and\ \bibinfo {author}
  {\bibfnamefont {C.}~\bibnamefont {Ciuti}},\ }\href@noop {} {} (\bibinfo
  {year} {2018}),\ \Eprint {http://arxiv.org/abs/1812.08582} {arXiv:1812.08582
  [quant-ph]} \BibitemShut {NoStop}%
\bibitem [{\citenamefont {LeCun}\ \emph {et~al.}(2015)\citenamefont {LeCun},
  \citenamefont {Bengio},\ and\ \citenamefont {Hinton}}]{LeCun2015}%
  \BibitemOpen
  \bibfield  {author} {\bibinfo {author} {\bibfnamefont {Y.}~\bibnamefont
  {LeCun}}, \bibinfo {author} {\bibfnamefont {Y.}~\bibnamefont {Bengio}}, \
  and\ \bibinfo {author} {\bibfnamefont {G.}~\bibnamefont {Hinton}},\ }\href
  {\doibase 10.1038/nature14539} {\bibfield  {journal} {\bibinfo  {journal}
  {Nature}\ }\textbf {\bibinfo {volume} {521}},\ \bibinfo {pages} {436}
  (\bibinfo {year} {2015})}\BibitemShut {NoStop}%
\bibitem [{\citenamefont {Carleo}\ and\ \citenamefont
  {Troyer}(2017)}]{CarleoScience2017}%
  \BibitemOpen
  \bibfield  {author} {\bibinfo {author} {\bibfnamefont {G.}~\bibnamefont
  {Carleo}}\ and\ \bibinfo {author} {\bibfnamefont {M.}~\bibnamefont
  {Troyer}},\ }\href {\doibase 10.1126/science.aag2302} {\bibfield  {journal}
  {\bibinfo  {journal} {Science}\ }\textbf {\bibinfo {volume} {355}},\ \bibinfo
  {pages} {602} (\bibinfo {year} {2017})}\BibitemShut {NoStop}%
\bibitem [{\citenamefont {Glasser}\ \emph {et~al.}(2018)\citenamefont
  {Glasser}, \citenamefont {Pancotti}, \citenamefont {August}, \citenamefont
  {Rodriguez},\ and\ \citenamefont {Cirac}}]{GlasserPRX2018}%
  \BibitemOpen
  \bibfield  {author} {\bibinfo {author} {\bibfnamefont {I.}~\bibnamefont
  {Glasser}}, \bibinfo {author} {\bibfnamefont {N.}~\bibnamefont {Pancotti}},
  \bibinfo {author} {\bibfnamefont {M.}~\bibnamefont {August}}, \bibinfo
  {author} {\bibfnamefont {I.~D.}\ \bibnamefont {Rodriguez}}, \ and\ \bibinfo
  {author} {\bibfnamefont {J.~I.}\ \bibnamefont {Cirac}},\ }\href {\doibase
  10.1103/physrevx.8.011006} {\bibfield  {journal} {\bibinfo  {journal} {Phys.
  Rev. X}\ }\textbf {\bibinfo {volume} {8}},\ \bibinfo {pages} {011006}
  (\bibinfo {year} {2018})}\BibitemShut {NoStop}%
\bibitem [{\citenamefont {van Nieuwenburg}\ \emph {et~al.}(2017)\citenamefont
  {van Nieuwenburg}, \citenamefont {Liu},\ and\ \citenamefont
  {Huber}}]{Nieuwenburg2017}%
  \BibitemOpen
  \bibfield  {author} {\bibinfo {author} {\bibfnamefont {E.~P.~L.}\
  \bibnamefont {van Nieuwenburg}}, \bibinfo {author} {\bibfnamefont {Y.-H.}\
  \bibnamefont {Liu}}, \ and\ \bibinfo {author} {\bibfnamefont {S.~D.}\
  \bibnamefont {Huber}},\ }\href {https://doi.org/10.1038/nphys4037} {\bibfield
   {journal} {\bibinfo  {journal} {Nature Physics}\ }\textbf {\bibinfo {volume}
  {13}},\ \bibinfo {pages} {435 EP } (\bibinfo {year} {2017})}\BibitemShut
  {NoStop}%
\bibitem [{\citenamefont {Schindler}\ \emph {et~al.}(2017)\citenamefont
  {Schindler}, \citenamefont {Regnault},\ and\ \citenamefont
  {Neupert}}]{Schindler2017}%
  \BibitemOpen
  \bibfield  {author} {\bibinfo {author} {\bibfnamefont {F.}~\bibnamefont
  {Schindler}}, \bibinfo {author} {\bibfnamefont {N.}~\bibnamefont {Regnault}},
  \ and\ \bibinfo {author} {\bibfnamefont {T.}~\bibnamefont {Neupert}},\ }\href
  {\doibase 10.1103/PhysRevB.95.245134} {\bibfield  {journal} {\bibinfo
  {journal} {Phys. Rev. B}\ }\textbf {\bibinfo {volume} {95}},\ \bibinfo
  {pages} {245134} (\bibinfo {year} {2017})}\BibitemShut {NoStop}%
\bibitem [{\citenamefont {Choo}\ \emph {et~al.}(2018)\citenamefont {Choo},
  \citenamefont {Carleo}, \citenamefont {Regnault},\ and\ \citenamefont
  {Neupert}}]{Choo2018}%
  \BibitemOpen
  \bibfield  {author} {\bibinfo {author} {\bibfnamefont {K.}~\bibnamefont
  {Choo}}, \bibinfo {author} {\bibfnamefont {G.}~\bibnamefont {Carleo}},
  \bibinfo {author} {\bibfnamefont {N.}~\bibnamefont {Regnault}}, \ and\
  \bibinfo {author} {\bibfnamefont {T.}~\bibnamefont {Neupert}},\ }\href
  {\doibase 10.1103/PhysRevLett.121.167204} {\bibfield  {journal} {\bibinfo
  {journal} {Phys. Rev. Lett.}\ }\textbf {\bibinfo {volume} {121}},\ \bibinfo
  {pages} {167204} (\bibinfo {year} {2018})}\BibitemShut {NoStop}%
\bibitem [{\citenamefont {Czischek}\ \emph {et~al.}(2018)\citenamefont
  {Czischek}, \citenamefont {G\"arttner},\ and\ \citenamefont
  {Gasenzer}}]{Czischek2018}%
  \BibitemOpen
  \bibfield  {author} {\bibinfo {author} {\bibfnamefont {S.}~\bibnamefont
  {Czischek}}, \bibinfo {author} {\bibfnamefont {M.}~\bibnamefont
  {G\"arttner}}, \ and\ \bibinfo {author} {\bibfnamefont {T.}~\bibnamefont
  {Gasenzer}},\ }\href {\doibase 10.1103/PhysRevB.98.024311} {\bibfield
  {journal} {\bibinfo  {journal} {Phys. Rev. B}\ }\textbf {\bibinfo {volume}
  {98}},\ \bibinfo {pages} {024311} (\bibinfo {year} {2018})}\BibitemShut
  {NoStop}%
\bibitem [{\citenamefont {Sharir}\ \emph {et~al.}(2019)\citenamefont {Sharir},
  \citenamefont {Levine}, \citenamefont {Wies}, \citenamefont {Carleo},\ and\
  \citenamefont {Shashua}}]{DirectSampling2019}%
  \BibitemOpen
  \bibfield  {author} {\bibinfo {author} {\bibfnamefont {O.}~\bibnamefont
  {Sharir}}, \bibinfo {author} {\bibfnamefont {Y.}~\bibnamefont {Levine}},
  \bibinfo {author} {\bibfnamefont {N.}~\bibnamefont {Wies}}, \bibinfo {author}
  {\bibfnamefont {G.}~\bibnamefont {Carleo}}, \ and\ \bibinfo {author}
  {\bibfnamefont {A.}~\bibnamefont {Shashua}},\ }\href@noop {} {\  (\bibinfo
  {year} {2019})},\ \Eprint {http://arxiv.org/abs/1902.04057} {arXiv:1902.04057
  [cond-mat.dis-nn]} \BibitemShut {NoStop}%
\bibitem [{\citenamefont {Daley}(2014)}]{Daley2014}%
  \BibitemOpen
  \bibfield  {author} {\bibinfo {author} {\bibfnamefont {A.~J.}\ \bibnamefont
  {Daley}},\ }\href {\doibase 10.1080/00018732.2014.933502} {\bibfield
  {journal} {\bibinfo  {journal} {Advances in Physics}\ }\textbf {\bibinfo
  {volume} {63}},\ \bibinfo {pages} {77} (\bibinfo {year} {2014})}\BibitemShut
  {NoStop}%
\bibitem [{\citenamefont {Albert}\ and\ \citenamefont
  {Jiang}(2014)}]{AlbertPRA14}%
  \BibitemOpen
  \bibfield  {author} {\bibinfo {author} {\bibfnamefont {V.~V.}\ \bibnamefont
  {Albert}}\ and\ \bibinfo {author} {\bibfnamefont {L.}~\bibnamefont {Jiang}},\
  }\href {\doibase 10.1103/PhysRevA.89.022118} {\bibfield  {journal} {\bibinfo
  {journal} {Phys. Rev. A}\ }\textbf {\bibinfo {volume} {89}},\ \bibinfo
  {pages} {022118} (\bibinfo {year} {2014})}\BibitemShut {NoStop}%
\bibitem [{\citenamefont {Spohn}(1977)}]{Spohn1977}%
  \BibitemOpen
  \bibfield  {author} {\bibinfo {author} {\bibfnamefont {H.}~\bibnamefont
  {Spohn}},\ }\href {\doibase 10.1007/BF00420668} {\bibfield  {journal}
  {\bibinfo  {journal} {Letters in Mathematical Physics}\ }\textbf {\bibinfo
  {volume} {2}},\ \bibinfo {pages} {33} (\bibinfo {year} {1977})}\BibitemShut
  {NoStop}%
\bibitem [{\citenamefont {Minganti}\ \emph {et~al.}(2018)\citenamefont
  {Minganti}, \citenamefont {Biella}, \citenamefont {Bartolo},\ and\
  \citenamefont {Ciuti}}]{PhysRevA.98.042118}%
  \BibitemOpen
  \bibfield  {author} {\bibinfo {author} {\bibfnamefont {F.}~\bibnamefont
  {Minganti}}, \bibinfo {author} {\bibfnamefont {A.}~\bibnamefont {Biella}},
  \bibinfo {author} {\bibfnamefont {N.}~\bibnamefont {Bartolo}}, \ and\
  \bibinfo {author} {\bibfnamefont {C.}~\bibnamefont {Ciuti}},\ }\href
  {\doibase 10.1103/PhysRevA.98.042118} {\bibfield  {journal} {\bibinfo
  {journal} {Phys. Rev. A}\ }\textbf {\bibinfo {volume} {98}},\ \bibinfo
  {pages} {042118} (\bibinfo {year} {2018})}\BibitemShut {NoStop}%
\bibitem [{\citenamefont {Prosen}(2014)}]{Prosen2014PRL}%
  \BibitemOpen
  \bibfield  {author} {\bibinfo {author} {\bibfnamefont {T.}~\bibnamefont
  {Prosen}},\ }\href {\doibase 10.1103/PhysRevLett.112.030603} {\bibfield
  {journal} {\bibinfo  {journal} {Phys. Rev. Lett.}\ }\textbf {\bibinfo
  {volume} {112}},\ \bibinfo {pages} {030603} (\bibinfo {year}
  {2014})}\BibitemShut {NoStop}%
\bibitem [{\citenamefont {Prosen}(2011)}]{Prosen2011PRL}%
  \BibitemOpen
  \bibfield  {author} {\bibinfo {author} {\bibfnamefont {T.}~\bibnamefont
  {Prosen}},\ }\href {\doibase 10.1103/PhysRevLett.107.137201} {\bibfield
  {journal} {\bibinfo  {journal} {Phys. Rev. Lett.}\ }\textbf {\bibinfo
  {volume} {107}},\ \bibinfo {pages} {137201} (\bibinfo {year}
  {2011})}\BibitemShut {NoStop}%
\bibitem [{\citenamefont {Torlai}\ and\ \citenamefont
  {Melko}(2018)}]{Torlai2018}%
  \BibitemOpen
  \bibfield  {author} {\bibinfo {author} {\bibfnamefont {G.}~\bibnamefont
  {Torlai}}\ and\ \bibinfo {author} {\bibfnamefont {R.~G.}\ \bibnamefont
  {Melko}},\ }\href {\doibase 10.1103/PhysRevLett.120.240503} {\bibfield
  {journal} {\bibinfo  {journal} {Phys. Rev. Lett.}\ }\textbf {\bibinfo
  {volume} {120}},\ \bibinfo {pages} {240503} (\bibinfo {year}
  {2018})}\BibitemShut {NoStop}%
\bibitem [{Note1()}]{Note1}%
  \BibitemOpen
  \bibinfo {note} {For more details, see Supplemental Material at the end of
  this PDF document}\BibitemShut {NoStop}%
\bibitem [{\citenamefont {Roux}\ and\ \citenamefont
  {Bengio}(2008)}]{LeRoux2008NeurComp}%
  \BibitemOpen
  \bibfield  {author} {\bibinfo {author} {\bibfnamefont {N.~L.}\ \bibnamefont
  {Roux}}\ and\ \bibinfo {author} {\bibfnamefont {Y.}~\bibnamefont {Bengio}},\
  }\href {\doibase 10.1162/neco.2008.04-07-510} {\bibfield  {journal} {\bibinfo
   {journal} {Neural Computation}\ }\textbf {\bibinfo {volume} {20}},\ \bibinfo
  {pages} {1631} (\bibinfo {year} {2008})}\BibitemShut {NoStop}%
\bibitem [{\citenamefont {Younes}(1996)}]{Younes1996AML}%
  \BibitemOpen
  \bibfield  {author} {\bibinfo {author} {\bibfnamefont {L.}~\bibnamefont
  {Younes}},\ }\href {\doibase 10.1016/0893-9659(96)00041-9} {\bibfield
  {journal} {\bibinfo  {journal} {Applied Mathematics Letters}\ }\textbf
  {\bibinfo {volume} {9}},\ \bibinfo {pages} {109} (\bibinfo {year}
  {1996})}\BibitemShut {NoStop}%
\bibitem [{\citenamefont {Montufar}\ and\ \citenamefont
  {Ay}(2011)}]{Montufar2011NeurComp}%
  \BibitemOpen
  \bibfield  {author} {\bibinfo {author} {\bibfnamefont {G.}~\bibnamefont
  {Montufar}}\ and\ \bibinfo {author} {\bibfnamefont {N.}~\bibnamefont {Ay}},\
  }\href {\doibase 10.1162/neco_a_00113} {\bibfield  {journal} {\bibinfo
  {journal} {Neural Computation}\ }\textbf {\bibinfo {volume} {23}},\ \bibinfo
  {pages} {1306} (\bibinfo {year} {2011})}\BibitemShut {NoStop}%
\bibitem [{\citenamefont {Weimer}(2015)}]{Weimer2015}%
  \BibitemOpen
  \bibfield  {author} {\bibinfo {author} {\bibfnamefont {H.}~\bibnamefont
  {Weimer}},\ }\href {\doibase 10.1103/physrevlett.114.040402} {\bibfield
  {journal} {\bibinfo  {journal} {Phys. Rev. Lett.}\ }\textbf {\bibinfo
  {volume} {114}},\ \bibinfo {pages} {040402} (\bibinfo {year}
  {2015})}\BibitemShut {NoStop}%
\bibitem [{Note2()}]{Note2}%
  \BibitemOpen
  \bibinfo {note} {Being able to rewrite the cost function (\ref
  {eq:l2-cost-function}) into this form allows us to sample it efficiently.
  This is the main advantage with respect to using the trace norm of $\protect
  \mathcal {L}\protect \mathaccentV {hat}05E{\rho }$ \cite {Weimer2015} as a
  cost function, which requires computing the singular-value-decomposition of
  $\protect \mathcal {L}\rho $ at each iteration.}\BibitemShut {Stop}%
\bibitem [{Note3()}]{Note3}%
  \BibitemOpen
  \bibinfo {note} {This is not the only possible expression that allows to
  sample the cost function $\protect \mathcal {C}(\protect \bm {v})$. However,
  this choice of the local cost function $\protect \mathcal {C}^{\protect \rm
  Loc}$ is particularly convenient since it respects the zero-variance
  property, it is more numerically stable and cheaper to compute. For further
  details see Sec. 3 of the Supplemental Material.}\BibitemShut {Stop}%
\bibitem [{\citenamefont {Becca}\ and\ \citenamefont
  {Sorella}(2017{\natexlab{a}})}]{Becca2017}%
  \BibitemOpen
  \bibfield  {author} {\bibinfo {author} {\bibfnamefont {F.}~\bibnamefont
  {Becca}}\ and\ \bibinfo {author} {\bibfnamefont {S.}~\bibnamefont
  {Sorella}},\ }\href {\doibase 10.1017/9781316417041} {\emph {\bibinfo {title}
  {Quantum Monte Carlo Approaches for Correlated Systems}}}\ (\bibinfo
  {publisher} {Cambridge University Press},\ \bibinfo {year}
  {2017})\BibitemShut {NoStop}%
\bibitem [{Note4()}]{Note4}%
  \BibitemOpen
  \bibinfo {note} {We point out that a promising direction of research would be
  to devise particular trial wave functions where $Z$ is fixed or cheaper to
  compute, so that a direct sampling of the distribution $p_{\sigma , \protect
  \mathaccentV {tilde}07E{\sigma }}$ without a Markov Chain would lead to
  easier convergence properties. Indeed, it has been recently shown \cite
  {DirectSampling2019} that a direct sampling is possible in some types of
  networks.}\BibitemShut {Stop}%
\bibitem [{\citenamefont {Bottou}(2010)}]{StochasticGradientDescent}%
  \BibitemOpen
  \bibfield  {author} {\bibinfo {author} {\bibfnamefont {L.}~\bibnamefont
  {Bottou}},\ }in\ \href@noop {} {\emph {\bibinfo {booktitle} {Proceedings of
  COMPSTAT'2010}}},\ \bibinfo {editor} {edited by\ \bibinfo {editor}
  {\bibfnamefont {Y.}~\bibnamefont {Lechevallier}}\ and\ \bibinfo {editor}
  {\bibfnamefont {G.}~\bibnamefont {Saporta}}}\ (\bibinfo  {publisher}
  {Physica-Verlag HD},\ \bibinfo {address} {Heidelberg},\ \bibinfo {year}
  {2010})\ pp.\ \bibinfo {pages} {177--186}\BibitemShut {NoStop}%
\bibitem [{\citenamefont {Becca}\ and\ \citenamefont
  {Sorella}(2017{\natexlab{b}})}]{BeccaBookChOptimization2017}%
  \BibitemOpen
  \bibfield  {author} {\bibinfo {author} {\bibfnamefont {F.}~\bibnamefont
  {Becca}}\ and\ \bibinfo {author} {\bibfnamefont {S.}~\bibnamefont
  {Sorella}},\ }in\ \href {\doibase 10.1017/9781316417041.007} {\emph {\bibinfo
  {booktitle} {Quantum Monte Carlo Approaches Correl. Syst.}}}\ (\bibinfo
  {publisher} {Cambridge University Press},\ \bibinfo {year} {2017})\ pp.\
  \bibinfo {pages} {131--155}\BibitemShut {NoStop}%
\bibitem [{\citenamefont {Jin}\ \emph {et~al.}(2018)\citenamefont {Jin},
  \citenamefont {Biella}, \citenamefont {Viyuela}, \citenamefont {Ciuti},
  \citenamefont {Fazio},\ and\ \citenamefont {Rossini}}]{Jin2018_ising}%
  \BibitemOpen
  \bibfield  {author} {\bibinfo {author} {\bibfnamefont {J.}~\bibnamefont
  {Jin}}, \bibinfo {author} {\bibfnamefont {A.}~\bibnamefont {Biella}},
  \bibinfo {author} {\bibfnamefont {O.}~\bibnamefont {Viyuela}}, \bibinfo
  {author} {\bibfnamefont {C.}~\bibnamefont {Ciuti}}, \bibinfo {author}
  {\bibfnamefont {R.}~\bibnamefont {Fazio}}, \ and\ \bibinfo {author}
  {\bibfnamefont {D.}~\bibnamefont {Rossini}},\ }\href {\doibase
  10.1103/PhysRevB.98.241108} {\bibfield  {journal} {\bibinfo  {journal} {Phys.
  Rev. B}\ }\textbf {\bibinfo {volume} {98}},\ \bibinfo {pages} {241108}
  (\bibinfo {year} {2018})}\BibitemShut {NoStop}%
\bibitem [{\citenamefont {Bezanson}\ \emph {et~al.}(2017)\citenamefont
  {Bezanson}, \citenamefont {Edelman}, \citenamefont {Karpinski},\ and\
  \citenamefont {Shah}}]{Julia2017}%
  \BibitemOpen
  \bibfield  {author} {\bibinfo {author} {\bibfnamefont {J.}~\bibnamefont
  {Bezanson}}, \bibinfo {author} {\bibfnamefont {A.}~\bibnamefont {Edelman}},
  \bibinfo {author} {\bibfnamefont {S.}~\bibnamefont {Karpinski}}, \ and\
  \bibinfo {author} {\bibfnamefont {V.~B.}\ \bibnamefont {Shah}},\ }\href
  {\doibase 10.1137/141000671} {\bibfield  {journal} {\bibinfo  {journal}
  {{SIAM} Review}\ }\textbf {\bibinfo {volume} {59}},\ \bibinfo {pages} {65}
  (\bibinfo {year} {2017})}\BibitemShut {NoStop}%
\bibitem [{\citenamefont {Kr\"{a}mer}\ \emph {et~al.}(2018)\citenamefont
  {Kr\"{a}mer}, \citenamefont {Plankensteiner}, \citenamefont {Ostermann},\
  and\ \citenamefont {Ritsch}}]{quantumopticsjl}%
  \BibitemOpen
  \bibfield  {author} {\bibinfo {author} {\bibfnamefont {S.}~\bibnamefont
  {Kr\"{a}mer}}, \bibinfo {author} {\bibfnamefont {D.}~\bibnamefont
  {Plankensteiner}}, \bibinfo {author} {\bibfnamefont {L.}~\bibnamefont
  {Ostermann}}, \ and\ \bibinfo {author} {\bibfnamefont {H.}~\bibnamefont
  {Ritsch}},\ }\href {\doibase 10.1016/j.cpc.2018.02.004} {\bibfield  {journal}
  {\bibinfo  {journal} {Computer Physics Communications}\ }\textbf {\bibinfo
  {volume} {227}},\ \bibinfo {pages} {109} (\bibinfo {year}
  {2018})}\BibitemShut {NoStop}%
\bibitem [{\citenamefont {Johansson}\ \emph {et~al.}(2012)\citenamefont
  {Johansson}, \citenamefont {Nation},\ and\ \citenamefont {Nori}}]{qutip2012}%
  \BibitemOpen
  \bibfield  {author} {\bibinfo {author} {\bibfnamefont {J.}~\bibnamefont
  {Johansson}}, \bibinfo {author} {\bibfnamefont {P.}~\bibnamefont {Nation}}, \
  and\ \bibinfo {author} {\bibfnamefont {F.}~\bibnamefont {Nori}},\ }\href
  {\doibase 10.1016/j.cpc.2012.02.021} {\bibfield  {journal} {\bibinfo
  {journal} {Computer Physics Communications}\ }\textbf {\bibinfo {volume}
  {183}},\ \bibinfo {pages} {1760} (\bibinfo {year} {2012})}\BibitemShut
  {NoStop}%
\bibitem [{\citenamefont {Johansson}\ \emph {et~al.}(2013)\citenamefont
  {Johansson}, \citenamefont {Nation},\ and\ \citenamefont {Nori}}]{qutip2013}%
  \BibitemOpen
  \bibfield  {author} {\bibinfo {author} {\bibfnamefont {J.}~\bibnamefont
  {Johansson}}, \bibinfo {author} {\bibfnamefont {P.}~\bibnamefont {Nation}}, \
  and\ \bibinfo {author} {\bibfnamefont {F.}~\bibnamefont {Nori}},\ }\href
  {\doibase 10.1016/j.cpc.2012.11.019} {\bibfield  {journal} {\bibinfo
  {journal} {Computer Physics Communications}\ }\textbf {\bibinfo {volume}
  {184}},\ \bibinfo {pages} {1234} (\bibinfo {year} {2013})}\BibitemShut
  {NoStop}%
\bibitem [{\citenamefont {{Hartmann}}\ and\ \citenamefont
  {{Carleo}}(2019)}]{Hartmann-2019arXiv190205131H}%
  \BibitemOpen
  \bibfield  {author} {\bibinfo {author} {\bibfnamefont {M.~J.}\ \bibnamefont
  {{Hartmann}}}\ and\ \bibinfo {author} {\bibfnamefont {G.}~\bibnamefont
  {{Carleo}}},\ }\href@noop {} {\  (\bibinfo {year} {2019})},\ \Eprint
  {http://arxiv.org/abs/1902.05131} {arXiv:1902.05131 [quant-ph]} \BibitemShut
  {NoStop}%
\bibitem [{\citenamefont {Yoshioka}\ and\ \citenamefont
  {Hamazaki}(2019)}]{giappi19arxiv}%
  \BibitemOpen
  \bibfield  {author} {\bibinfo {author} {\bibfnamefont {N.}~\bibnamefont
  {Yoshioka}}\ and\ \bibinfo {author} {\bibfnamefont {R.}~\bibnamefont
  {Hamazaki}},\ }\href@noop {} {} (\bibinfo {year} {2019}),\ \Eprint
  {http://arxiv.org/abs/1902.07006} {arXiv:1902.07006 [cond-mat.dis-nn]}
  \BibitemShut {NoStop}%
\bibitem [{\citenamefont {Nagy}\ and\ \citenamefont
  {Savona}(2019)}]{Nagy19arxiv}%
  \BibitemOpen
  \bibfield  {author} {\bibinfo {author} {\bibfnamefont {A.}~\bibnamefont
  {Nagy}}\ and\ \bibinfo {author} {\bibfnamefont {V.}~\bibnamefont {Savona}},\
  }\href@noop {} {} (\bibinfo {year} {2019}),\ \Eprint
  {http://arxiv.org/abs/1902.09483} {arXiv:1902.09483 [quant-ph]} \BibitemShut
  {NoStop}%
\end{thebibliography}%
